\documentclass[11pt]{article}
\usepackage{verbatim,amsmath,amssymb}
\usepackage{epsfig,float,color}
\usepackage{geometry}
\usepackage{setspace}
\usepackage{natbib}
\usepackage{amsthm,mathrsfs,amsfonts,dsfont} 
\usepackage{hyperref}

\definecolor{darkblue}{rgb}{0,0,1}
\definecolor{col1}{rgb}{1,0,1}
\definecolor{col2}{rgb}{0.1,.75,0}

\hypersetup{pdftex=true, colorlinks=true, breaklinks=true, linkcolor=darkblue, menucolor=darkblue, pagecolor=darkblue, citecolor=darkblue, urlcolor=darkblue}

\newcommand{\bitm}{\begin{itemize}}
\newcommand{\eitm}{\end{itemize}}
\newcommand{\bnumr}{\begin{enumerate}}
\newcommand{\enumr}{\end{enumerate}}

\newcommand{\mrT}{\mathrm{T}}

\newcommand {\eqb}[1]{\begin{equation}\begin{array}{#1}}
\newcommand {\eqe}{\end{array}\end{equation}}

\newcommand {\esb}[1]{\begin{equation*}\begin{array}{#1}}
\newcommand {\ese}{\end{array}\end{equation*}}
\newcommand {\ds}{\displaystyle}

\newcommand {\pa}[2]{\frac{\partial{#1}}{\partial{#2}}}
\newcommand {\pad}[2]{\frac{\mathrm{d}{#1}}{\mathrm{d}{#2}}}

\newcommand {\back}{\! \! \!}
\newcommand {\is}{\back &=& \back}
\newcommand {\dis}{\back &:=& \back}

\newcommand {\plus}{\back &+& \back}
\newcommand {\mi}{\back &-& \back}

\newcommand {\isleq}{\back &\leq& \back}

\newcommand {\isgeq}{\back &\geq& \back}

\newcommand {\norm}[1]{\|#1\|}
\newcommand {\tr}{\mathrm{tr}\,}

\newcommand {\grad}{\mathrm{grad}\,}

\newcommand {\dif}{\mathrm{d}}

\newcommand {\II}{{I\kern-.3em I}}
\newcommand {\III}{{I\kern-.3em I\kern-.3em I}}

\newcommand {\mrb}{\mathrm{b}}
\newcommand {\mrc}{\mathrm{c}}

\newcommand {\mre}{\mathrm{e}}

\newcommand {\mrg}{\mathrm{g}}

\newcommand {\mri}{\mathrm{i}}

\newcommand {\mrm}{\mathrm{m}}

\newcommand {\mrs}{\mathrm{s}}

\newcommand {\mry}{\mathrm{y}}

\newcommand {\ba}{\boldsymbol{a}}
\newcommand {\bb}{\boldsymbol{b}}

\newcommand {\be}{\boldsymbol{e}}
\newcommand {\bff}{\boldsymbol{f}}

\newcommand {\bi}{\boldsymbol{i}}
\newcommand {\bj}{\boldsymbol{j}}

\newcommand {\bm}{\boldsymbol{m}}
\newcommand {\bn}{\boldsymbol{n}}

\newcommand {\bq}{\boldsymbol{q}}

\newcommand {\bv}{\boldsymbol{v}}

\newcommand {\bx}{\boldsymbol{x}}

\newcommand {\bmu}{\boldsymbol{\mu}}

\newcommand {\bnu}{\mbox{\boldmath$\nu$}}

\newcommand {\bA}{\boldsymbol{A}}
\newcommand {\bB}{\boldsymbol{B}}
\newcommand {\bC}{\boldsymbol{C}}
\newcommand {\bD}{\boldsymbol{D}}
\newcommand {\bE}{\boldsymbol{E}}
\newcommand {\bF}{\boldsymbol{F}}

\newcommand {\bI}{\boldsymbol{I}}

\newcommand {\bM}{\boldsymbol{M}}
\newcommand {\bN}{\boldsymbol{N}}

\newcommand {\bS}{\boldsymbol{S}}
\newcommand {\bT}{\boldsymbol{T}}

\newcommand {\bV}{\boldsymbol{V}}

\newcommand {\bX}{\boldsymbol{X}}

\newcommand {\eps}{\varepsilon}
\newcommand {\sig}{\sigma}

\newcommand {\bsig}{\mbox{\boldmath$\sigma$}}

\newcommand {\bone}{\mathbf{1}}

\newcommand {\IR}{{\rm\kern.24em
   \vrule width.02em height1.53ex depth-.05ex
   \kern-.3em R}}
\newcommand {\ic}{{\rm\kern.20em
   \vrule width.02em height1.0ex depth-.05ex
   \kern-.22em c}}
\newcommand {\ia}{{\rm\kern.20em
   \vrule width.02em height1.05ex depth-.0ex
   \kern-.25em a}}
\newcommand {\IC}{{\rm\kern.24em
   \vrule width.02em height1.4ex depth-.05ex
   \kern-.26em C}}
\newcommand {\ID}{{\rm\kern.34em
   \vrule width.02em height1.5ex depth-.05ex
   \kern-.36em D}}
\newcommand {\IS}{{\rm\kern.24em
   \vrule width.02em height1.6ex depth.05ex
   \kern-.26em S}}
\newcommand {\IT}{{\rm\kern.50em
   \vrule width.02em height1.55ex depth-.05ex
   \kern-.52em T}}

\newcommand {\IE}{{\rm\kern.24em
   \vrule width.02em height1.55ex depth-.05ex
   \kern-.33em E}}
\newcommand {\IEa}{{\rm\kern.24em
   \vrule width.02em height1.55ex depth-.05ex
   \kern-.33em E}^{1}_{ijkl}}
\newcommand {\IEb}{{\rm\kern.24em
   \vrule width.02em height1.55ex depth-.05ex
   \kern-.33em E}^{2}_{ijkl}}

\newcommand {\sB}{\mathcal{B}}

\newcommand {\sM}{\mathcal{M}}

\newcommand {\sP}{\mathcal{P}}

\newcommand {\sR}{\mathcal{R}}
\newcommand {\sS}{\mathcal{S}}

\newcommand {\Ass}[2]{\kern 0.9ex \vrule width0.45em height0.2ex depth0ex \kern -2.1ex \bigwedge_{#1}^{#2}}
\newcommand {\ASS}[2]{\kern 1.45ex \vrule width0.5em height0.2ex depth0ex \kern -2.65ex \bigwedge_{#1}^{#2}}

\newcommand{\el}{\mathrm{el}}
\newcommand{\inel}{\mathrm{in}}

\begin{document}

\begin{center}
\Large{\bf{The multiplicative deformation split for shells with application to growth, chemical swelling, thermoelasticity, viscoelasticity and elastoplasticity}}\\

\end{center}

\begin{center}
\large{Roger A.~Sauer$^{\ast,}$\footnote{corresponding author, email: sauer@aices.rwth-aachen.de}, Reza Ghaffari$^\ast$ and Anurag Gupta$^\dag$} \\
\vspace{4mm}

\small{\textit{$^\ast$Aachen Institute for Advanced Study in Computational Engineering Science (AICES), \\ 
RWTH Aachen University, Templergraben 55, 52056 Aachen, Germany \\[1mm]
$^\dag$Department of Mechanical Engineering, Indian Institute of Technology Kanpur, UP 208016, India}}

\vspace{4mm}

Published\footnote{This pdf is the personal version of an article whose final publication is available at \href{http://dx.doi.org/10.1016/j.ijsolstr.2019.06.002}{www.sciencedirect.com}} 
in \textit{Int.~J.~Solids Struc.}, 
\href{http://dx.doi.org/10.1016/j.ijsolstr.2019.06.002}{DOI: 10.1016/j.ijsolstr.2019.06.002} \\
Submitted on 3.~December 2018, Revised on 20.~May 2019, Accepted on 4.~June 2019 

\end{center}

\vspace{3mm}


\rule{\linewidth}{.15mm}
{\bf Abstract}

This work presents a general unified theory for coupled nonlinear elastic and inelastic deformations of curved thin shells.
The coupling is based on a multiplicative decomposition of the surface deformation gradient.
The kinematics of this decomposition is examined in detail.
In particular, the dependency of various kinematical quantities, such as area change and curvature, on the elastic and inelastic strains is discussed. 
This is essential for the development of general constitutive models.
In order to fully explore the coupling between elastic and different inelastic deformations, the surface balance laws for mass, momentum, energy and entropy are examined in the context of the multiplicative decomposition.
Based on the second law of thermodynamics, the general constitutive relations are then derived.
Two cases are considered: Independent inelastic strains, and inelastic strains that are functions of temperature and concentration.
The constitutive relations are illustrated by several nonlinear examples on growth, chemical swelling, thermoelasticity, viscoelasticity and elastoplasticity of shells. 
The formulation is fully expressed in curvilinear coordinates leading to compact and elegant 
expressions for the kinematics, balance laws and constitutive relations. 


\vspace{-5mm}
\rule{\linewidth}{.15mm}

\section{Introduction}\label{s:intro}

Many problems in science and technology are characterized by different, competing deformation types.
Apart from elastic deformations, which are studied predominantly in solid mechanics, deformations can also arise from growth, swelling, thermal expansion, viscosity, plasticity and electro-magnetical fields. 
The decomposition of these deformations is essential for the proper modeling and understanding of coupled problems.
In thermoelasticity, for instance, mechanical stresses do not arise from thermal deformations, but from the elastic deformations countering those. 
In the general framework of large deformations, the decomposition of deformations is based on the multiplicative split of the deformation gradient.
While the topic has been studied extensively for three-dimensional continua, there are much fewer works studying the multiplicative split for curved surfaces. 
In particular, a general shell theory that unifies different deformation types is currently lacking and therefore addressed here.

The origins of the multiplicative decomposition of the deformation gradient can be traced back to \citet{Flory50}, who used a 1D version of it to decompose elastic and inelastic stretches during swelling, \citet{Eckart48} and \citet{Kondo49}, who introduced the notion of a locally relaxed (stress-free) intermediate configuration that can be globally incompatible, and \citet{Bilby57} and \citet{Kroner59}, who formalized the multiplicative decomposition for plasticity. 
Recent discussion on the origin, mathematical nature and application of the multiplicative decomposition has been provided by \citet{Lubarda04}, \citet{gupta07} and \citet{reina18}. 
Following its introduction for swelling and plasticity, the multiplicative decomposition has been extended to thermoelasticity \citep{Stojanovic64}, viscoelasticity \citep{Sidoroff74} and growth \citep{Kondaurov87,Takamizawa87}. 
Subsequently, a vast literature body has appeared on the topic.
Most of it deals with 3D continua or shell formulations derived from those using the \textit{degenerate solid} framework \citep{ahmad70,Parisch78}.
These cases are based on a deformation decomposition in 3D -- usually in the context of Cartesian coordinate systems.
Instead of this, we are concerned here with a decomposition of the surface deformation in the general framework of curvilinear coordinates.
Therefore we restrict the following survey to general shell structures based on such surface formulations. 

\textbf{Growth and swelling of shells:}
The first general surface formulations using a multiplicative decomposition to couple mechanical deformation and growth seem to be the works by \citet{Dervaux09b} and \citet{Wang18} on plates, \citet{Rausch14} on membranes, \citet{Vetter13,Vetter14} on Kirchhoff-Love shells, and \citet{Lychev14} on Reissner-Mindlin shells.  
Similar approaches have also been considered by \citet{Papastavrou13} to model surface growth of bulk materials and \citet{swain18} to model interface growth.
A general surface formulation coupling mechanical deformation and swelling of membranes and shells seems to have appeared only recently by the work of \citet{Lucantonio17}.

\textbf{Viscoelasticity of shells:}
The work by \citet{Neff05a} seems to be the first general surface model for viscoelastic shells and membranes that is using a multiplicative decomposition of the deformation. 
All subsequent works seem to have resorted back to additive decompositions.
Examples are the works by \citet{Lubarda11} on erythrocyte membranes, \citet{Li12b} on the derivation of shell formulations from 3D viscoelasticity, \citet{Altenbach15} on micropolar shells and \citet{Dorr17} on fiber reinforced composite shells.

\textbf{Elastoplasticity of shells:}
The research on general elastoplastic shells goes back to the works by \citet{Green68}, \citet{Sawczuk82}, \citet{basar91} and \citet{simo92b}.
They follow however an additive decomposition of the strain.
The work by \citet{simo92b}, seems to be the first FE model that is directly based on a surface formulation instead of considering the thickness integration of 3D continua, as has been done by many others, e.g.~see \citet{Stumpf94}, \citet{Miehe98}, \citet{Betsch99a}, \citet{Eberlein99} and recently \citet{steigmann15} and \citet{ambati18}.
This property clearly distinguishes these works from the approach taken here: 
Instead of integrating 3D continua, here the entire theory, including the multiplicative decomposition, is directly based on a surface formulation. 
The recent surface formulations by \citet{Dujc12} and \citet{roychowdhury18}, on the other hand, are based again on an additive split,
although the existence of a multiplicative split of the surface deformation gradient has been alluded to in the latter work.

\textbf{Thermomechanics of shells:}
General surface formulations for thermomechanical shells have been developed by \citet{Green79}, \citet{Reddy98} and recently \citet{Kar16}.
However, none of these formulations is based on a multiplicative decomposition.
Instead, a general surface formulation for thermomechanical shells based on a multiplicative decomposition seems to be still lacking.

The survey shows that even though many works have appeared on coupled elastic and inelastic deformations for shells, only few use a general surface formulation in the general framework of curvilinear coordinates, and none seem to start from a multiplicative decomposition of the surface deformation gradient. 
Instead they either start from an additive decomposition or they are based on thickness integration of the classical multiplicative decomposition.

The use of a curvilinear coordinate description allows for a very general treatment of shell geometry and deformation.
Also it allows for a direct finite element (FE) formulation that avoids the overhead of a transformation to a Cartesian formulation as is classically used. 
Shell FE formulations tend to be much more efficient that classical 3D FE formulations, since no thickness discretization is needed.
Instead, simplifying assumptions are used for the thickness behavior.  
The most efficient shell formulation is based on Kirchhoff-Love kinematics, which assumes that cross-sections remain planar and normal to the mid-plane during deformation.
Further, a normal thickness stress is usually neglected.
These assumptions are suitable for thin shells, whose planar dimensions are at least one order of magnitude larger than the thickness.\\
Following the works by \citet{prigogine}, \citet{degroot}, \citet{naghdi72} and \citet{steigmann99} on irreversible thermodynamics and nonlinear Kirchhoff-Love shell theory, \citet{shelltheo} and \citet{sahu17} recently developed a new multiphysical shell theory that is suitable for both solid and liquid shells. 
Based on this theory, new finite element formulations have been proposed for engineering shells \citep{solidshell}, layered shells \citep{composhell}, biological shells \citep{bioshell}, graphene \citep{graphene}, lipid bilayers \citep{liquidshell}, inverse analysis \citep{inverseshell}, phase transformations \citep{phaseshell} and surfactants \citep{surfactant}.
However, all of these works are restricted to elastic deformations.

The restrictions in the current literature mentioned above motivates the development of a general thin shell formulation for coupled deformations.
Such a formulation should be based on a multiplicative split, in order to handle large deformations, use curvilinear coordinates, in order to handle general surface geometries, and account for the laws of irreversible thermodynamics, in order to capture the full scope of coupling. 
Compared to existing formulations, the formulation proposed here has several novelties: 
\begin{itemize}
\item It provides a unified shell theory for coupled nonlinear elastic and inelastic deformations.
\item It is based on the multiplicative decomposition of the surface deformation gradient.
\item It accounts for growth, swelling, viscosity, plasticity and thermal deformations.
\item It is fully formulated in the general and compact framework of curvilinear coordinates.
\item It explores the coupling in the kinematic relations and local balance laws.
\item It is illustrated by several constitutive examples derived from the second law of thermodynamics.
\end{itemize}
It is also shown that the multiplicative split on the surface deformation gradient generally leads to an additive split on certain strain components. 
Additive decompositions are therefore not restricted to small deformations.
Further, some of the existing formulations found in the literature are recovered as special cases of the proposed multiplicative split.

The remainder of this paper is organized as follows. 
Sec.~\ref{s:surf} gives a brief overview of the general, curvilinear description of curved surfaces.
Secs.~\ref{s:kin} and \ref{s:motion} then discuss the kinematics and motion of curved surfaces accounting for a multiplicative decomposition of the surface deformation gradient.
This is followed by the presentation of the balance laws of mass, momentum, energy and entropy in Sec.~\ref{s:bal}.
These lead to the coupled strong form problem statement, summarized in Sec.~\ref{s:prob}, and the general constitutive equations, derived in Sec.~\ref{s:consti}.
The latter are illustrated by several examples for elastic and inelastic material behavior.
Sec.~\ref{s:conc} concludes the paper.

\section{Mathematical surface description}\label{s:surf}

\begin{figure}[h]
\begin{center} \unitlength1cm
\begin{picture}(0,7)
\put(-6.2,-.3){\includegraphics[height=76mm]{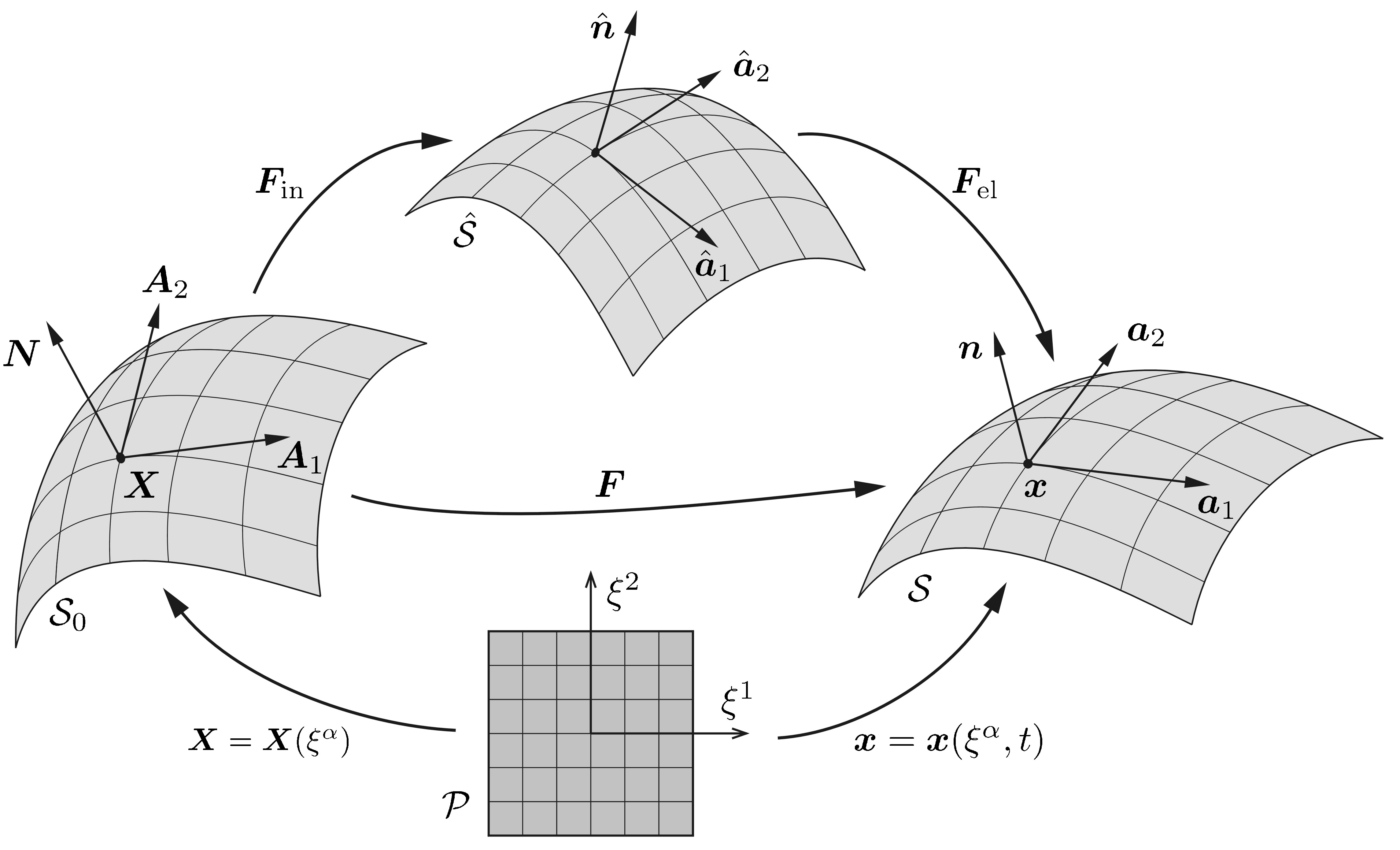}}
\end{picture}
\caption{Surface description and kinematics: Multiplicative split into elastic and inelastic deformations and their relation to reference, intermediate and current configurations $\sS_0$, $\hat\sS$ and~$\sS$.}
\label{f:split}
\end{center}
\end{figure}

This section gives a brief summary of the three-dimensional mathematical description of curved surfaces in the general framework of curvilinear coordinates.
It allows to track any surface deformation (discussed in Sec.~\ref{s:kin}), and it is particularly suited for subsequent finite element formulations.
In this framework, every point $\bx$ on a surface $\sS$ is given by the mapping
\eqb{l}
\bx = \bx(\xi^\alpha,t)\,.
\label{e:bx}\eqe
Here $\xi^\alpha$, for $\alpha=1,2$, denotes the two curvilinear coordinates that can be associated with a 2D parameter domain $\sP$ as illustrated in Fig.~\ref{f:split}, while $t\in[0,t_\mathrm{end}]$ stands for time.
Given \eqref{e:bx}, all geometrical aspects of the surface can be obtained.
The tangent plane at $\bx\in\sS$ is characterized by the two tangent vectors 
\eqb{l}
\ba_\alpha := \ds\pa{\bx}{\xi^\alpha}\,,
\eqe
while the surface normal at $\bx\in\sS$ is given by 
\eqb{l}
\bn = \ds\frac{\ba_1\times\ba_2}{\norm{\ba_1\times\ba_2}}\,.
\label{e:bn}\eqe
The basis $\ba_1$, $\ba_2$ and $\bn$ allows to introduce the notion of in-plane and out-of-plane surface objects.
The tangent vectors $\ba_1$ and $\ba_2$ are generally not orthonormal, meaning that the so-called surface metric,
\eqb{l}
a_{\alpha\beta} := \ba_\alpha\cdot\ba_\beta\,,
\label{e:a_ab}\eqe
generally gives $[a_{\alpha\beta}] \neq [1~0\,; 0~1]$.
To restore orthonormality a set of dual tangent vectors $\ba^\alpha$ is introduced from $\ba_\alpha = a_{\alpha\beta}\,\ba^\beta$ and $\ba^\alpha = a^{\alpha\beta}\ba_\beta$,\footnote{Following index notation, summation (from 1 to 2) is implied on all terms with repeated Greek indices, i.e.~$a^{\alpha\beta}\ba_\beta =  a^{\alpha1}\ba_1 + a^{\alpha2}\ba_2$.} 
where $[a^{\alpha\beta}] := [a_{\alpha\beta}]^{-1}$, such that $\ba^\alpha\cdot\ba_\beta=\delta^\alpha_\beta$ for $[\delta^\alpha_\beta] = [1~0\,; 0~1]$. 
This illustrates a very important property of $a_{\alpha\beta}$ and $a^{\alpha\beta}$: 
They lower or raise indices.
\\
Another important surface characteristic is the curvature.
It follows from the out-of-plane components of the second derivative $\ba_{\alpha,\beta}:=\partial\ba_\alpha/\partial\xi^\beta$, denoted as
\eqb{l}
b_{\alpha\beta} := \ba_{\alpha,\beta}\cdot\bn\,.
\label{e:b_ab}\eqe
These components can be arranged in the matrix $[b^\alpha_{~\beta}] := [a^{\alpha\gamma}b_{\gamma\beta}]$, whose eigenvalues are the two principal surface curvatures
\eqb{l}
\kappa_\alpha = H \pm \sqrt{H^2-\kappa} \,,
\label{e:kap_a}\eqe
where
\eqb{lll}
H \dis \frac{1}{2}\,a^{\alpha\beta}\,b_{\alpha\beta} \\[1mm]
\kappa \dis \det[b^\alpha_{~\beta}]\\
\label{e:Hkappa}\eqe
are the mean curvature and Gaussian curvature of surface $\sS$, respectively.
The derivative $\ba_{\alpha,\beta}$ is also referred to as the parametric derivate of $\ba_\alpha$.
It is generally different to the so-called \textit{co-variant derivative} of $\ba_\alpha$ that is denoted by ``;" and defined as
\eqb{l}
\ba_{\alpha;\beta} := (\bn\otimes\bn)\,\ba_{\alpha,\beta}\,.
\eqe  
For general scalars and vectors (that have no free index) the parametric and co-variant derivates are identical.
Only for objects with free indices (such as $\ba_\alpha$ and $\ba^\alpha$) a difference appears.  
\\
Analogous to \eqref{e:bx}, physical fields on $\sS$ are generally functions of $\xi^\alpha$ and t.
Examples are surface density $\rho=\rho(\xi^\alpha,t)$ and surface temperature $T=T(\xi^\alpha,t)$.
Their surface gradient follows from $\rho_{,\alpha}(=\rho_{;\alpha})$ and $T_{,\alpha}(=T_{;\alpha})$ as 
$\grad_{\!\mrs}\rho=\rho_{,\alpha}\,\ba^\alpha$ and $\grad_{\!\mrs} T=T_{,\alpha}\,\ba^\alpha$.
\\
A more comprehensive treatment of the mathematical description of curved surfaces can be found in the classical textbooks on differential geometry, e.g.~see \citet{Kreyszig}.
A recent concise treatment is also provided in \citet{cism}.

\section{Surface kinematics}\label{s:kin}

This section introduces reference, current and intermediate configuration, and discusses the kinematical quantities between them.
The discussion is restricted to Kirchhoff-Love kinematics.
These are entirely based on the notion of surface strain and curvature, and do not need any further kinematical measures.
The description follows the classical treatment found in the works by \citet{naghdi72,naghdi82}, \citet{pietraszkiewicz89} and \citet{libai}.

\subsection{Classical kinematical measures}\label{s:kin_c}

Suppose that the surface $\sS$ deforms over time.
The initial configuration at time $t=0$ is defined as reference configuration, and denoted $\sS_0$ to distinguish it from the current configuration $\sS$ at time $t>0$.
In order to distinguish all the surface quantities introduced in Sec.~\ref{s:surf}, upper case symbols (or the subscript ``0") are used for the reference configuration, while lower case symbols (or no subscript) are used for the current configuration, see Fig.~\ref{f:split}.
\\
The primary measure relating $\sS$ and $\sS_0$ is the surface deformation gradient
\eqb{l}
\bF := \ba_\alpha \otimes \bA^\alpha\,.
\eqe
Together with its generalized inverse
\eqb{l}
\bF^{-1} = \bA_\alpha \otimes \ba^\alpha\,,
\eqe
it transforms the tangent vectors as
\eqb{lll}
\ba_\alpha \is \bF\bA_\alpha\,, \\[1mm]
\bA_\alpha \is \bF^{-1}\ba_\alpha\,, \\[1mm]
\bA^\alpha \is \bF^\mrT\ba^\alpha\,, \\[1mm]
\ba^\alpha \is \bF^{-\mrT}\bA^\alpha\,.
\eqe
From $\bF$ follow the two surface Cauchy Green tensors
\eqb{lllll}
\bC \dis \bF^\mrT\bF \is a_{\alpha\beta}\,\bA^\alpha\otimes\bA^\beta\,, \\[1mm]
\bB \dis \bF\bF^\mrT \is A^{\alpha\beta}\,\ba_\alpha\otimes\ba_\beta\,.
\eqe
From these, the surface Green-Lagrange strain tensor
\eqb{l}
\bE := \frac{1}{2}\big( \bC - \bI \big)
\label{e:E}\eqe
and the surface Almansi strain tensor
\eqb{l}
\be := \frac{1}{2}\big( \bi - \bB^{-1} \big)\,,
\label{e:e}\eqe
can be defined.
Here,
\eqb{lllll}
\bI \dis \bA_\alpha \otimes \bA^\alpha \is A_{\alpha\beta}\,\bA^\alpha\otimes\bA^\beta\,, \\[1mm]
\bi \dis \ba_\alpha \otimes \ba^\alpha \is a_{\alpha\beta}\,\ba^\alpha\otimes\ba^\beta
\label{e:bIi}\eqe
denote the surface identity tensors on $\sS_0$ and $\sS$, respectively.
$\bE$ has the components
\eqb{l}
E_{\alpha\beta} := \bA_\alpha\cdot\bE\bA_\beta = \frac{1}{2}\big( a_{\alpha\beta} - A_{\alpha\beta} \big)\,,
\eqe
w.r.t.~basis $\bA^\alpha$, while $\be$ has the components
\eqb{l}
e_{\alpha\beta} := \ba_\alpha\cdot\be\,\ba_\beta = \frac{1}{2}\big( a_{\alpha\beta} - A_{\alpha\beta} \big)
\eqe
w.r.t.~basis $\ba^\alpha$.
To emphasize the equality $e_{\alpha\beta}=E_{\alpha\beta}$ and, as is seen later, the fact that the multiplicative split on $\bF$ leads to an additive split on these strain components, we introduce
\eqb{l}
\eps_{\alpha\beta} := \frac{1}{2}\big( a_{\alpha\beta} - A_{\alpha\beta} \big)\,,
\label{e:eps}\eqe
such that
\eqb{lll}
\bE \is \eps_{\alpha\beta}\, \bA^\alpha\otimes\bA^\beta\,, \\[1mm]
\be \is \eps_{\alpha\beta}\, \ba^\alpha\otimes\ba^\beta\,.
\label{e:Ee}\eqe
Similar to \eqref{e:eps}, we introduce the relative curvature components
\eqb{l}
\kappa_{\alpha\beta} := b_{\alpha\beta} - B_{\alpha\beta}\,.
\label{e:kap_ab}\eqe
The surface Cauchy-Green tensors have two invariants, $I_1$ and $J$.
In order to define them, the surface determinant of $\bF$ is introduced by
\eqb{l}
\det_\mrs\bF := \ds\frac{\norm{\bF\,\bV_{\!1}\times\bF\,\bV_{\!2}}}{\norm{\bV_{\!1}\times\bV_{\!2}}}
\eqe
for all non-parallel surface tangent vectors $\bV_{\!1}$ and $\bV_{\!2}$ \citep{javili14}.
Picking $\bV_{\!\alpha}=\bA_\alpha$, this leads to the second invariant
\eqb{l}
J := \det_\mrs\bF = \ds\frac{\norm{\ba_1\times\ba_2}}{\norm{\bA_1\times\bA_2}}\,,
\label{e:Ja}\eqe
which is equal to
\eqb{l}
J = \ds\frac{\sqrt{\det{[a_{\alpha\beta}]}}}{\sqrt{\det{[A_{\alpha\beta}]}}}\,,
\label{e:J}\eqe
and corresponds to the local change of area between $\sS_0$ and $\sS$.
The first invariant is
\eqb{l}
I_1 = \bI:\bC = \bi:\bB = A^{\alpha\beta}a_{\alpha\beta}\,.
\label{e:I1}\eqe

\subsection{Kinematics of the multiplicative deformation split}\label{s:kin_s}

The previous setting accounts only for a single deformation source.
Its primary unknown is position $\bx$, from which everything else follows.
In order to extend the setting to deformations composed of two separate (i.e.~elastic and inelastic) components, we introduce the intermediate surface configuration $\hat\sS$ with the tangent vectors $\hat\ba_\alpha$ that are now an additional set of unknowns.
The deformation $\sS_0\rightarrow\hat\sS$ is taken as the inelastic part, while $\hat\sS\rightarrow\sS$ is the elastic part, see Fig.~\ref{f:split}.
Given the tangent vectors $\hat\ba_\alpha$, the surface normal $\hat\bn$, metric $\hat a_{\alpha\beta}$, inverse metric $\hat a^{\alpha\beta}$, dual tangent vectors $\hat\ba^\alpha$, curvature components $\hat b_{\alpha\beta}$, mean curvature $\hat H$ and Gaussian curvature $\hat\kappa$ are obtained analogous to expressions \eqref{e:bn}-\eqref{e:Hkappa} in Sec.~\ref{s:surf}. 
The introduced intermediate configuration $\hat\sS$ implies that the surface deformation gradient $\bF$ can be multiplicatively split as
\eqb{l}
\bF = \bF_{\!\el}\, \bF_{\!\inel}\,,
\label{e:FeFi}\eqe
where
\eqb{lll}
\bF_{\!\el} \dis \ba_\alpha \otimes \hat\ba^\alpha\,, \\[1mm] 
\bF_{\!\inel} \dis \hat\ba_\alpha \otimes \bA^\alpha
\eqe
are the elastic and inelastic surface deformation gradients, respectively.
Split \eqref{e:FeFi} implies the inverse split
\eqb{l}
\bF^{-1} = \bF^{-1}_{\!\inel}\,\bF^{-1}_{\!\el}\,,
\eqe
with
\eqb{lll}
\bF_{\!\el}^{-1} \dis \hat\ba_\alpha \otimes \ba^\alpha\,, \\[1mm] 
\bF_{\!\inel}^{-1} \dis \bA_\alpha \otimes \hat\ba^\alpha\,.
\eqe
$\bF_{\!\el}$ and $\bF_{\!\inel}$ transform the tangent vectors as
\eqb{lll}
\ba_\alpha \is \bF_{\!\el}\,\hat\ba_\alpha\,, \\[1mm]
\hat\ba_\alpha \is \bF^{-1}_{\!\el}\ba_\alpha\,, \\[1mm]
\hat\ba^\alpha \is \bF^\mrT_{\!\el}\ba^\alpha\,, \\[1mm]
\ba^\alpha \is \bF^{-\mrT}_{\!\el}\hat\ba^\alpha
\eqe
and
\eqb{lll}
\hat\ba_\alpha \is \bF_{\!\inel}\,\bA_\alpha\,, \\[1mm]
\bA_\alpha \is \bF^{-1}_{\!\inel}\hat\ba_\alpha\,, \\[1mm]
\bA^\alpha \is \bF^\mrT_{\!\inel}\hat\ba^\alpha\,, \\[1mm]
\hat\ba^\alpha \is \bF^{-\mrT}_{\!\inel}\bA^\alpha\,.
\eqe
These relations can be used to push forward the right surface Cauchy-Green tensor $\bC$ to the intermediate configuration, i.e.
\eqb{l}
\bC_\el := \bF^{-\mrT}_{\!\inel}\bC\,\bF^{-1}_{\!\inel} = \bF^\mrT_{\!\el}\bF_{\!\el} = a_{\alpha\beta}\,\hat\ba^\alpha\otimes\hat\ba^\beta\,,
\eqe
and to pull back the inverse left Cauchy-Green tensor $\bB^{-1}$ to the intermediate configuration, i.e.
\eqb{l}
\bB^{-1}_\inel := \bF^\mrT_{\!\el}\bB^{-1}\bF_{\!\el} = \bF^{-\mrT}_{\!\inel}\bF^{-1}_{\!\inel} = A_{\alpha\beta}\,\hat\ba^\alpha\otimes\hat\ba^\beta\,.
\eqe
In order to decompose the strain, it is convenient to introduce the strain tensor
\eqb{l}
\hat\be := \eps_{\alpha\beta}\,\hat\ba^\alpha\otimes\hat\ba^\beta
\eqe
analogous to \eqref{e:Ee}.
This strain corresponds to the push forward of $\bE$ and the pull-back of $\be$ to the intermediate configuration, since
\eqb{l}
\hat\be = \bF^{-\mrT}_{\!\inel}\bE\,\bF^{-1}_{\!\inel} = \bF^\mrT_{\!\el}\,\be\,\bF_{\!\el}\,.
\label{e:hate}\eqe
Inserting \eqref{e:E} or \eqref{e:e} into \eqref{e:hate} gives
\eqb{l}
\hat\be = \frac{1}{2}\big(\bC_\el - \bB^{-1}_\inel \big)
\eqe
which admits the simple additive decomposition
\eqb{l}
\hat\be = \hat\be_\el + \hat\be_\inel
\eqe
based on the definitions
\eqb{lll}
\hat\be_\el \dis \frac{1}{2}\big(\bC_\el - \hat\bi \big) \\[1.5mm]
\hat\be_\inel \dis \frac{1}{2}\big(\hat\bi - \bB^{-1}_\inel \big)
\eqe
that are analogous to \eqref{e:E} and \eqref{e:e}. 
Here,
\eqb{l}
\hat\bi := \hat\ba_\alpha \otimes \hat\ba^\alpha = \hat a_{\alpha\beta}\,\hat\ba^\alpha\otimes\hat\ba^\beta
\eqe
denotes the surface identity on $\hat\sS$ analogous to \eqref{e:bIi}.
Introducing
\eqb{lll}
\hat\be_\el \is \eps_{\alpha\beta}^\el\,\hat\ba^\alpha\otimes\hat\ba^\beta\,, \\[1mm]
\hat\be_\inel \is \eps_{\alpha\beta}^\inel\,\hat\ba^\alpha\otimes\hat\ba^\beta
\eqe
then leads to the componentwise decomposition
\eqb{l}
\eps_{\alpha\beta} = \eps_{\alpha\beta}^\el + \eps_{\alpha\beta}^\inel\,,
\label{e:eps2}\eqe
with
\eqb{lll}
\eps_{\alpha\beta}^\el \is \frac{1}{2}\big(a_{\alpha\beta}-\hat a_{\alpha\beta}\big)\,, \\[1.5mm]  
\eps_{\alpha\beta}^\inel \is \frac{1}{2}\big(\hat a_{\alpha\beta}-A_{\alpha\beta}\big)\,,
\label{e:eps_ei}\eqe
which is analogous to \eqref{e:eps}.
Thus we have showed that the multiplicative decomposition of $\bF$ generally leads to the additive decomposition of the intermediate strain tensor $\hat\be$.
An alternative, but less insightful approach is to directly propose an additive decomposition of the strain components without introducing $\bF=\bF_{\!\el}\bF_{\!\inel}$, as has for example been done 
by \citet{Reddy98} for thermoelastic shells,
by \citet{simo92b} for elastoplastic shells,
by \citet{Lubarda11} for viscoelastic shells,
by \citet{Liang11} for shell growth,
by \citet{Sman15} for shell swelling, and
by \citet{roychowdhury18} for surface defects.
In the more general context of 3D continua, the additive decomposition \eqref{e:eps2}-\eqref{e:eps_ei} goes back to \citet{sedov}
\\
We note that the proposed multiplicative split, apart from being more general than an additive split, equally applies to growth, swelling, viscosity, plasticity and thermal deformations, and thus allows for their unified treatment.

Likewise to decomposition \eqref{e:eps2}, we introduce the additive curvature decomposition
\eqb{l}
\kappa_{\alpha\beta} = \kappa_{\alpha\beta}^\el + \kappa_{\alpha\beta}^\inel\,,
\label{e:kap2}\eqe
with
\eqb{lll}
\kappa_{\alpha\beta}^\el \dis b_{\alpha\beta}-\hat b_{\alpha\beta}\,, \\[1.5mm]  
\kappa_{\alpha\beta}^\inel \dis \hat b_{\alpha\beta}-B_{\alpha\beta}\,.
\label{e:kap_ei}\eqe

Due to split \eqref{e:FeFi}, the local area change, introduced in \eqref{e:Ja}-\eqref{e:J}, becomes
\eqb{l}
J = J_\el\,J_\inel\,,
\label{e:JeJi}\eqe
with
\eqb{lllll}
J_\el \is \det_\mrs\bF_{\!\el} = \ds\frac{\norm{\ba_1\times\ba_2}}{\norm{\hat\ba_1\times\hat\ba_2}} 
	\is \ds\frac{\sqrt{\det{[a_{\alpha\beta}]}}}{\sqrt{\det{[\hat a_{\alpha\beta}]}}}\,, \\[4mm]
J_\inel \is \det_\mrs\bF_{\!\inel} = \ds\frac{\norm{\hat\ba_1\times\hat\ba_2}}{\norm{\bA_1\times\bA_2}} 
	\is \ds\frac{\sqrt{\det{[\hat a_{\alpha\beta}]}}}{\sqrt{\det{[A_{\alpha\beta}]}}}\,.
\label{e:Jei}\eqe

A further useful object is the first invariant of $\bC_\el$, given by the surface trace
\eqb{l}
I^\el_1 := \tr_{\!\mrs}\,\bC_\el = \bC_\el:\hat\bi = \hat a^{\alpha\beta}a_{\alpha\beta}\,.
\label{e:I1e}\eqe
It is equivalent to the surface trace $\tr_{\!\mrs}\,\bB_\el=\bB_\el:\bi$, where $\bB_\el = \bF_{\!\el}\,\bF_{\!\el}^\mrT$.\footnote{Here, the surface trace of a surface tensor is defined w.r.t.~the surface the tensor refers to.}
Likewise the first invariant of $\bC_\el^{-1}$ is given by
\eqb{l}
I_{1-}^\el := \tr_{\!\mrs}\,\bC^{-1}_\el = \bC^{-1}_\el:\hat\bi =  a^{\alpha\beta} \hat a_{\alpha\beta} = \tr_{\!\mrs}\,\bB^{-1}_\el =\bB^{-1}_\el:\bi \,.
\label{e:I1e_inv}\eqe
Note, that in general $I_{1-}^\el\neq 1/I_1^\el$.

\textbf{Remark 1:} The inelastic deformation does not need to be compatible, i.e.~$\bF_{\!\inel}$ does not need to follow as the gradient of a deformation mapping.
Instead it can be treated as an independent unknown. 
For a recent discussion on incompatible plastic deformations, see \citet{gupta07}.

\textbf{Remark 2:} A mutiplicative decomposition can also be used to split the elastic deformation into two parts, e.g.~a pre-strain and an additional strain. 
In this case, the intermediate configuration $\hat\sS$ is not stress-free but (pre-)stressed.
The total stress then depends on the total strain in the usual way, and so the kinematical decomposition discussed above is not needed.

\subsection{Inelastic dilatation}\label{s:indil}

For many applications, like growth or thermal expansion, the inelastic deformation is purely dilatational.
Excluding rigid body rotations (which can be accounted for in the elastic deformation), inelastic dilatation is described by the intermediate tangent vectors
\eqb{l}
\hat\ba_\alpha = \lambda_\inel\,\bA_\alpha\,,
\label{e:hatba_dil}\eqe
such that
\eqb{l}
\bF_{\!\inel} = \lambda_\inel\,\bI\,.
\label{e:Fin_dil}\eqe
Here $\lambda_\inel = \sqrt{J_\inel}$ denotes the inelastic stretch.
As a consequence,
\eqb{lll}
\hat a_{\alpha\beta} \is J_\inel\,A_{\alpha\beta}\,, \\[1mm]
\hat a^{\alpha\beta} \is A^{\alpha\beta}/J_\inel
\label{e:hata_dil}\eqe 
and $\hat\ba^\alpha = \bA^\alpha/\lambda_\inel$.
From this follows
\eqb{l}
I_1^\el = I_1/J_\inel\,.
\eqe

\subsection{Inelastic isotropic bending}

Analogous to inelastic dilatation is the case of inelastic isotropic bending.
In this case, the two principal curvatures defined in \eqref{e:kap_a} increase by the same scalar factor $\bar\kappa_\inel$ during inelastic deformation, i.e.~the intermediate configuration is characterized by the principal curvatures
\eqb{l}
\hat\kappa_\alpha = \bar\kappa_\inel\,\kappa_{0\alpha}\,,
\label{e:hatk_dil}\eqe
where $\kappa_{0\alpha}$ are the two principal curvatures of $\sS_0$.
This implies
\eqb{l}
\hat H = \bar\kappa_\inel\,H_0\,,\quad
\hat\kappa = \bar\kappa_\inel^2\,\kappa_0\,.
\eqe
This follows for example from considering the curvature relation
\eqb{l}
\hat b^\alpha_{~\beta} = \bar\kappa_\inel\,B^\alpha_{~\beta}\,,
\eqe
which, together with \eqref{e:hata_dil}, implies
\eqb{lll}
\hat b_{\alpha\beta} = J_\inel\,\bar\kappa_\inel\,B_{\alpha\beta}\,, \\[1mm]
\hat b^{\alpha\beta} = \bar\kappa_\inel\,B^{\alpha\beta}/J_\inel\,.
\label{e:hatb_dil}\eqe

\section{Surface motion}\label{s:motion}

The kinematical quantities introduced in the preceding section generally change over time, which is discussed in this section.
To characterize these changes, we introduce the material time derivative
\eqb{l}
\dot{(...)} := \ds\pad{...}{t} := \ds\pa{...}{t}\bigg|_{\bX\,=\,\mathrm{fixed}}\,.
\label{e:dot}\eqe
From \eqref{e:bx} thus follows the surface velocity $\bv:=\dot\bx$.
In the following two subsections, we summarize some of the consequences of \eqref{e:dot} for the classical kinematical measures introduced in Sec.~\ref{s:kin_c} and the kinematical decomposition of Sec.~\ref{s:kin_s}. 
While the expressions in Sec.~\ref{s:motion_c} appear in older works, those of Sec.~\ref{s:motion_s} are mostly new.
They are required for formulating general constitutive models, as is discussed later (Sec.~\ref{s:consti}).
Sec.~\ref{s:motion_s} also exposes the coupling of elastic and inelastic contributions that are present in some of the kinematical quantities.

\subsection{Classical measures of the surface motion}\label{s:motion_c}

From definition \eqref{e:eps} follows the strain rate
\eqb{l}
\dot\eps_{\alpha\beta} = \frac{1}{2}\dot a_{\alpha\beta}\,,
\label{e:deps}\eqe
where
\eqb{l}
\dot a_{\alpha\beta} = \dot\ba_\alpha\cdot\ba_\beta + \ba_\alpha\cdot\dot\ba_\beta\,,
\label{e:daab}\eqe
according to \eqref{e:a_ab}.
With this we can find the material time derivative of the area change $J$.
Since $J = J(a_{\alpha\beta}) = J(\eps_{\alpha\beta})$,\footnote{To simplify notation, the same symbol (here $J$) is used for the variable and its different functions.}
\eqb{l}
\dot J = \ds\pa{J}{a_{\alpha\beta}}\,\dot a_{\alpha\beta} = \ds\pa{J}{\eps_{\alpha\beta}}\,\dot\eps_{\alpha\beta}
\label{e:dJ}\eqe
From \eqref{e:J} and \eqref{e:deps} follows
\eqb{l}
\ds\pa{J}{\eps_{\alpha\beta}} = 2\ds\pa{J}{a_{\alpha\beta}} = J\,a^{\alpha\beta}\,,
\eqe
so that
\eqb{l}
\ds\frac{\dot J}{J} = \ds\frac{1}{2}\,a^{\alpha\beta}\,\dot a_{\alpha\beta}\,.
\label{e:dJ}\eqe
Similarly, the first invariant of $\bC$, $I_1 = I_1(a_{\alpha\beta}) = I_1(\eps_{\alpha\beta})$, gives
\eqb{l}
\dot I_1 = \ds\pa{I_1}{a_{\alpha\beta}}\,\dot a_{\alpha\beta} = \ds\pa{I_1}{\eps_{\alpha\beta}}\,\dot\eps_{\alpha\beta}\,,
\eqe
where
\eqb{l}
\ds\pa{I_1}{\eps_{\alpha\beta}} = 2\pa{I_1}{a_{\alpha\beta}} = 2\,A^{\alpha\beta}\,,
\eqe
due to \eqref{e:I1} and \eqref{e:deps}.
\\
In order to express various curvature rates, we require
\eqb{l}
\dot\bn = - \big(\ba^\alpha\otimes\bn\big)\,\dot\ba_\alpha\,,
\label{e:dn}\eqe
and
\eqb{l}
\dot a^{\alpha\beta} = \ds\pa{a^{\alpha\beta}}{a_{\gamma\delta}} \dot a_{\gamma\delta}\,,
\label{e:daIab}\eqe
with
\eqb{l}
a^{\alpha\beta\gamma\delta} := \ds\pa{a^{\alpha\beta}}{a_{\gamma\delta}} = -\frac{1}{2}\big(a^{\alpha\gamma}a^{\beta\delta} + a^{\alpha\delta}a^{\beta\gamma}\big)\,,
\label{e:aIabcd}\eqe
see \citet{cism}.
From \eqref{e:kap_ab} then follows the relative curvature rate\
\eqb{l}
\dot\kappa_{\alpha\beta} = \dot b_{\alpha\beta}\,,
\label{e:dkap}\eqe
with
\eqb{l}
\dot b_{\alpha\beta} = \dot\ba_{\alpha,\beta}\cdot\bn + \ba_{\alpha,\beta}\cdot\dot\bn\,,
\label{e:db_ab}\eqe
due to \eqref{e:b_ab}.
Further, since $H = H(a_{\alpha\beta},a_{\alpha\beta})$ and  $\kappa = \kappa(a_{\alpha\beta},a_{\alpha\beta})$, we find the mean curvature rate
\eqb{l}
\dot H = \ds\pa{H}{a_{\alpha\beta}}\,\dot a_{\alpha\beta} + \pa{H}{b_{\alpha\beta}}\,\dot b_{\alpha\beta}\,,
\label{e:dH}\eqe
with
\eqb{l}
\ds\pa{H}{a_{\alpha\beta}} = -\frac{1}{2}\,b^{\alpha\beta}\,, \quad
\ds\pa{H}{b_{\alpha\beta}} = \frac{1}{2}\,a^{\alpha\beta}\,,
\label{e:dH2}\eqe
and the Gaussian curvature rate
\eqb{l}
\dot\kappa = \ds\pa{\kappa}{a_{\alpha\beta}}\,\dot a_{\alpha\beta} + \pa{\kappa}{b_{\alpha\beta}}\,\dot b_{\alpha\beta}\,,
\label{e:dkappa1}\eqe
with
\eqb{l}
\ds\pa{\kappa}{a_{\alpha\beta}} = -\kappa\,a^{\alpha\beta}\,, \quad
\ds\pa{\kappa}{b_{\alpha\beta}} = 2H\,a^{\alpha\beta}-b^{\alpha\beta}\,,
\label{e:dkappa}\eqe
due to \eqref{e:Hkappa} and \eqref{e:daIab}; see also \citet{cism}.
Relations \eqref{e:deps} and \eqref{e:dkap} obviously imply
\eqb{l}
\ds\pa{...}{\eps_{\alpha\beta}} = 2\pa{...}{a_{\alpha\beta}}\,,\quad
\ds\pa{...}{\kappa_{\alpha\beta}} = \pa{...}{b_{\alpha\beta}}\,.
\label{d:eps}\eqe

\subsection{Decomposition of the surface motion}\label{s:motion_s}

The additive strain decomposition of \eqref{e:eps2} and \eqref{e:eps_ei} directly leads to the additive rate decomposition
\eqb{lll}
\dot\eps_{\alpha\beta} \is \dot\eps_{\alpha\beta}^\el + \dot\eps_{\alpha\beta}^\inel\,, \\[1.5mm]
\dot\eps_{\alpha\beta}^\el \is \frac{1}{2}\big(\dot a_{\alpha\beta}-\dot{\hat a}_{\alpha\beta}\big)\,, \\[1.5mm]
\dot\eps_{\alpha\beta}^\inel \is \frac{1}{2}\dot{\hat a}_{\alpha\beta}\,,
\label{e:deps2}\eqe
where
\eqb{l}
\dot{\hat a}_{\alpha\beta} = \dot{\hat\ba}_\alpha\cdot\hat\ba_\beta + \hat\ba_\alpha\cdot\dot{\hat\ba}_\beta\,.
\label{e:dhaab}\eqe
Also the multiplicative decomposition of $J$ leads to an additive rate decomposition:
From \eqref{e:JeJi} directly follows
\eqb{l}
\ds\frac{\dot J}{J} = \frac{\dot J_\el}{J_\el} + \frac{\dot J_\inel}{J_\inel}\,.
\label{e:dJ2}\eqe
In order to determine $\dot J_\el$ and $\dot J_\inel$, we first note that for a general function $f(a_{\alpha\beta},\hat a_{\alpha\beta}) = f(\eps^\el_{\alpha\beta},\eps^\inel_{\alpha\beta})$ we have\footnote{Again the same symbol (here $f$) is used for the variable and its different functions.}
\eqb{l}
\dot f = \ds\pa{f}{a_{\alpha\beta}}\,\dot a_{\alpha\beta} + \pa{f}{\hat a_{\alpha\beta}}\,\dot{\hat a}_{\alpha\beta} 
= \ds\pa{f}{\eps^\el_{\alpha\beta}}\,\dot\eps^\el_{\alpha\beta} + \pa{f}{\eps^\inel_{\alpha\beta}}\,\dot\eps^\inel_{\alpha\beta}\,.
\label{e:dphi}\eqe
From \eqref{e:deps2} then follows
\eqb{l}
\ds\pa{f}{\eps^\el_{\alpha\beta}} = 2\ds\pa{f}{a_{\alpha\beta}}\,,\quad 
\ds\pa{f}{\eps^\inel_{\alpha\beta}} = 2\ds\pa{f}{a_{\alpha\beta}} + 2\pa{f}{\hat a_{\alpha\beta}}\,.
\label{d:phi}\eqe
Combing this with \eqref{d:eps} leads to
\eqb{l}
\ds\pa{f}{\eps^\el_{\alpha\beta}} = \pa{f}{\eps_{\alpha\beta}}\,.
\label{d:eps2}\eqe
Applying \eqref{e:dphi} and \eqref{d:phi} to $f= J_\inel(\hat a_{\alpha\beta}) = J_\inel(\eps^\inel_{\alpha\beta})$ defined in (\ref{e:Jei}.2) gives
\eqb{l}
\ds\pa{J_\inel}{\eps^\el_{\alpha\beta}} = 0\,,\quad 
\ds\pa{J_\inel}{\eps^\inel_{\alpha\beta}} = J_\inel\,\hat a^{\alpha\beta}
\eqe
and
\eqb{l}
\ds\frac{\dot J_\inel}{J_\inel} = \ds\frac{1}{2}\,\hat a^{\alpha\beta}\,\dot{\hat a}_{\alpha\beta}\,.
\label{e:dJin2}\eqe
Applying \eqref{e:dphi}  and \eqref{d:phi} to $f= J_\el(a_{\alpha\beta},\hat a_{\alpha\beta}) = J_\el(\eps^\el_{\alpha\beta},\eps^\inel_{\alpha\beta})$ defined in (\ref{e:Jei}.1) gives
\eqb{l}
\ds\pa{J_\el}{\eps^\el_{\alpha\beta}} = J_\el\,a^{\alpha\beta}\,,\quad
\ds\pa{J_\el}{\eps^\inel_{\alpha\beta}} = J_\el\,\big(a^{\alpha\beta} - \hat a^{\alpha\beta}\big)
\label{d:Jel_Eel}\eqe
and
\eqb{l}
\ds\frac{\dot J_\el}{J_\el} = \ds\frac{1}{2}\,a^{\alpha\beta}\,\dot a_{\alpha\beta} - \frac{1}{2}\,\hat a^{\alpha\beta}\,\dot{\hat a}_{\alpha\beta}\,.
\label{e:dJel2}\eqe
The later equation agrees with \eqref{e:dJ}, \eqref{e:dJ2} and \eqref{e:dJin2}.
\\
Applying \eqref{d:phi} to $f=I_1^\el = I_1^\el(a_{\alpha\beta},\hat a_{\alpha\beta}) = I_1^\el(\eps^\el_{\alpha\beta},\eps^\inel_{\alpha\beta})$ defined in \eqref{e:I1e} gives
\eqb{l}
\ds\pa{I_1^\el}{\eps^\el_{\alpha\beta}} = 2\,\hat a^{\alpha\beta}\,,\quad
\ds\pa{I_1^\el}{\eps^\inel_{\alpha\beta}} =  2\,\hat a^{\alpha\beta\gamma\delta}\big(a_{\gamma\delta}-\hat a_{\gamma\delta}\big)\,,
\label{d:Iel_Eel}\eqe
where
\eqb{l}
\hat a^{\alpha\beta\gamma\delta} := \ds\pa{\hat a^{\alpha\beta}}{\hat a_{\gamma\delta}} = -\frac{1}{2}\big(\hat a^{\alpha\gamma}\hat a^{\beta\delta} + \hat a^{\alpha\delta}\hat a^{\beta\gamma}\big)\,,
\label{e:haIabcd}\eqe
analogous to
\eqref{e:aIabcd}.
$\dot I_1^\el$ can then be obtained from \eqref{e:dphi}.
\\
Next we turn towards the curvature rates.
The additive curvature decomposition in \eqref{e:kap2} and \eqref{e:kap_ei} leads to the additive rate decomposition
\eqb{lll}
\dot\kappa_{\alpha\beta} \is \dot\kappa_{\alpha\beta}^\el + \dot\kappa_{\alpha\beta}^\inel\,, \\[1.5mm]
\dot\kappa_{\alpha\beta}^\el \is \dot b_{\alpha\beta}-\dot{\hat b}_{\alpha\beta}\,, \\[1.5mm]
\dot\kappa_{\alpha\beta}^\inel \is \dot{\hat b}_{\alpha\beta}\,,
\label{e:dkap2}\eqe
where
\eqb{l}
\dot{\hat b}_{\alpha\beta} = \dot{\hat\ba}_{\alpha,\beta}\cdot\hat\bn + \hat\ba_{\alpha,\beta}\cdot\dot{\hat\bn}\,,
\eqe
and
\eqb{l}
\dot{\hat\bn} = - \big(\hat\ba^\alpha\otimes\hat\bn\big)\,\dot{\hat\ba}_\alpha\,,
\eqe
analogous to \eqref{e:dn} and \eqref{e:db_ab}. 
In order to find various curvature rates, we first note that for a general function 
$f\big(a_{\alpha\beta},\hat a_{\alpha\beta},b_{\alpha\beta},\hat b_{\alpha\beta}\big) = f\big(\eps^\el_{\alpha\beta},\eps^\inel_{\alpha\beta},\kappa^\el_{\alpha\beta},\kappa^\inel_{\alpha\beta}\big)$ we have
\eqb{lllllllll}
\dot f 
\is \ds\pa{f}{a_{\alpha\beta}}\,\dot a_{\alpha\beta} \plus \ds\pa{f}{\hat a_{\alpha\beta}}\,\dot{\hat a}_{\alpha\beta} \plus \ds\pa{f}{b_{\alpha\beta}}\,\dot b_{\alpha\beta} \plus \ds\pa{f}{\hat b_{\alpha\beta}}\,\dot{\hat b}_{\alpha\beta}\,, \\[4.5mm]
\is \ds\pa{f}{\eps^\el_{\alpha\beta}}\,\dot\eps^\el_{\alpha\beta} \plus \ds\pa{f}{\eps^\inel_{\alpha\beta}}\,\dot\eps^\inel_{\alpha\beta} \plus \ds\pa{f}{\kappa^\el_{\alpha\beta}}\,\dot\kappa^\el_{\alpha\beta} \plus \ds\pa{f}{\kappa^\inel_{\alpha\beta}}\,\dot\kappa^\inel_{\alpha\beta}\,.
\label{e:dphi2}\eqe
From \eqref{e:deps2} and \eqref{e:dkap2} then follow
\eqb{l}
\ds\pa{f}{\kappa^\el_{\alpha\beta}} = \ds\pa{f}{b_{\alpha\beta}}\,,\quad 
\ds\pa{f}{\kappa^\inel_{\alpha\beta}} = \ds\pa{f}{b_{\alpha\beta}} + \pa{f}{\hat b_{\alpha\beta}}
\label{d:phi2}\eqe
together with the already known expressions in \eqref{d:phi}.
Combing this with \eqref{d:eps} further leads to
\eqb{l}
\ds\pa{f}{\kappa^\el_{\alpha\beta}} = \pa{f}{\kappa_{\alpha\beta}}\,. 
\label{d:kap2}\eqe
Applying \eqref{d:phi} and \eqref{d:phi2} to $f = \hat H(\hat a_{\alpha\beta},\hat b_{\alpha\beta})= \hat H(\eps^\inel_{\alpha\beta},\kappa^\inel_{\alpha\beta})$ defined by $\hat H = \hat a^{\alpha\beta}\hat b_{\alpha\beta}/2$ gives
\eqb{l}
\ds\pa{\hat H}{\eps^\el_{\alpha\beta}} = 0\,,\quad \ds\pa{\hat H}{\eps^\inel_{\alpha\beta}} = -\hat b^{\alpha\beta}\,, \quad
\ds\pa{\hat H}{\kappa^\el_{\alpha\beta}} = 0\,,\quad \ds\pa{\hat H}{\kappa^\inel_{\alpha\beta}} = \frac{1}{2}\,\hat a^{\alpha\beta}\,,
\label{e:dhH}\eqe
analogous to \eqref{e:dH2}.
$\dot{\hat H}$ can then be obtained from \eqref{e:dphi2}.\\
Applying \eqref{d:phi} and \eqref{d:phi2} to $f = \hat\kappa(\hat a_{\alpha\beta},\hat b_{\alpha\beta})=\hat\kappa(\eps^\inel_{\alpha\beta},\kappa^\inel_{\alpha\beta})$ defined by $\hat\kappa = \det[\hat b_{\alpha\beta}]/\det[\hat a_{\alpha\beta}]$ gives
\eqb{l}
\ds\pa{\hat\kappa}{\eps^\el_{\alpha\beta}} = 0\,,\quad \ds\pa{\hat\kappa}{\eps^\inel_{\alpha\beta}} = -2\hat\kappa\,\hat a^{\alpha\beta}\,, \quad
\ds\pa{\hat\kappa}{\kappa^\el_{\alpha\beta}} = 0\,,\quad \ds\pa{\hat\kappa}{\kappa^\inel_{\alpha\beta}} = 2\hat H\,\hat a^{\alpha\beta}-\hat b^{\alpha\beta}\,,
\label{e:dhk}\eqe
analogous to \eqref{e:dkappa}.
$\dot{\hat\kappa}$ can then be obtained from \eqref{e:dphi2}.
The fact that the elastic derivatives in \eqref{e:dhH} and \eqref{e:dhk} are zero underlines the fact that the intermediate configuration is an independent unknown that is independent of $\eps^\el_{\alpha\beta}$ and $\kappa^\el_{\alpha\beta}$.
\\
On top of those expressions, the constitutive models discussed in Sec.~\ref{s:consti} require the dependency of $H$ and $\kappa$ on the elastic and inelastic strain rates.
Applying \eqref{d:phi} and \eqref{d:phi2} to $f = H(a_{\alpha\beta},b_{\alpha\beta})=H(\eps^\el_{\alpha\beta},\eps^\inel_{\alpha\beta},\kappa^\el_{\alpha\beta},\kappa^\inel_{\alpha\beta})$ defined by \eqref{e:Hkappa}
gives
\eqb{l}
\ds\pa{H}{\eps^\el_{\alpha\beta}} = \ds\pa{H}{\eps^\inel_{\alpha\beta}} = -b^{\alpha\beta}\,, \quad
\ds\pa{H}{\kappa^\el_{\alpha\beta}} = \ds\pa{H}{\kappa^\inel_{\alpha\beta}} = \frac{1}{2}\,a^{\alpha\beta}\,,
\label{e:dHk}\eqe
due to \eqref{e:dH2}.
Applying \eqref{d:phi} and \eqref{d:phi2} to $f = \kappa(a_{\alpha\beta},b_{\alpha\beta})=\kappa(\eps^\el_{\alpha\beta},\eps^\inel_{\alpha\beta},\kappa^\el_{\alpha\beta},\kappa^\inel_{\alpha\beta})$ defined by \eqref{e:Hkappa}
gives
\eqb{l}
\ds\pa{\kappa}{\eps^\el_{\alpha\beta}} = \ds\pa{\kappa}{\eps^\inel_{\alpha\beta}} = -2\kappa\, a^{\alpha\beta}\,, \quad
\ds\pa{\kappa}{\kappa^\el_{\alpha\beta}} = \ds\pa{\kappa}{\kappa^\inel_{\alpha\beta}} = 2H\,a^{\alpha\beta} - b^{\alpha\beta}\,,
\label{e:dHk2}\eqe
due to \eqref{e:dkappa}.

\section{Surface balance laws}\label{s:bal}

This section discusses the balance laws of mass, momentum, energy and entropy for curved surfaces.
The derivation follows the framework of \citet{shelltheo} and \citet{sahu17}, which is based on the works by \citet{prigogine}, \citet{degroot}, \citet{naghdi72} and \citet{steigmann99}.
It makes use of three important theorems:
Reynold's transport theorem,
\eqb{l}
\ds\pad{}{t}\int_\sS...\,\dif a = \int_\sS\bigg(\dot{(...)} + \frac{\dot J}{J}\,(...) \bigg)\,\dif a\,,
\label{e:Rey}\eqe
which follows from substituting $\dif a = J\,\dif A$ and using the product rule, the surface divergence theorem,
\eqb{l}
\ds\int_{\partial\sS}...\,\nu_\alpha\,\dif s = \int_\sS ..._{;\alpha}\,\dif a\,,
\label{e:divtheo}\eqe
where $\nu_\alpha = \ba_\alpha\cdot\bnu$ is the in-plane component of the boundary normal $\bnu$ and "$;\alpha$" is the co-variant derivative defined in Sec.~\ref{s:surf}, and the localization theorem
\eqb{l}
\ds\int_\sR ...\,\dif a = 0 \quad \forall\,\sR\subset\sS \quad\Leftrightarrow\quad ... = 0 \quad\forall\,\bx\in\sS\,.
\label{e:localize}\eqe

\subsection{Surface mass balance}

Mass balance is formulated here for the case of a mixture of two species.
This could for example be a solvent diffusing into a matrix material and induce swelling.
The partial surface densities $\rho_1=\rho_1(\xi^\alpha,t)$ and $\rho_2=\rho_2(\xi^\alpha,t)$ are introduced such that the current surface density of the mixture is $\rho = \rho_1 + \rho_2$ (with unit mass per current area).

\subsubsection{Total mass balance}

Consider the Lagrangian description of the total mass balance
\eqb{l}
\ds\pad{}{t}\int_{\sR} \rho\,\dif a = \int_\sR h\,\dif a \qquad \forall\,\sR\subset\sS\,,
\eqe
where $h$ is a surface mass source (mass per current area and time) that originates for example from growth or swelling.
Applying Reynolds' transport theorem \eqref{e:Rey} and localization \eqref{e:localize} gives
\eqb{l}
\dot\rho + \rho\,\dot J/J = h \qquad \forall\,\bx\in\sS\,,
\label{e:ODE-rho}\eqe
which is the governing ODE for $\rho$.
In order to determine $\rho(t)$, the initial condition $\rho = \rho_0$ at $t=0$ is required.
\\
If $h=0$, then $\rho=\rho_0/J$ solves ODE \eqref{e:ODE-rho}, which elegantly eliminates unknown $\rho$ and its ODE. \\
If $h\neq0$, then ODE \eqref{e:ODE-rho} needs to be solved (numerically).
$h\neq0$ induces growth, which is an inelastic deformation. 
An example is isotropic growth discussed in Sec.~\ref{s:growth}. 

\subsubsection{Partial mass balance}

Additionally, the mass balance of the individual species needs to be accounted for. 
Given \eqref{e:ODE-rho}, it suffices to account for the mass balance of one species.
We therefore introduce the relative concentration $\phi = \rho_1/\rho$ of species 1 and denote its source term $h_1$.
The partial mass balance thus is
\eqb{l}
\ds\pad{}{t}\int_{\sR} \rho\,\phi\,\dif a = \int_\sR h_1\,\dif a + \int_{\partial\sR}j_\nu\,\dif s  \qquad \forall\,\sR\subset\sS \,,
\eqe
where the $j_\nu$ term accounts for a relative mass flux of species 1 w.r.t.~the average motion of the mixture. 
Defining the surface mass flux vector $\bj=j^\alpha\ba_\alpha$ through
\eqb{l}
j_\nu = - \bj\cdot\bnu = -j^\alpha\,\nu_\alpha
\eqe
and using the surface divergence theorem \eqref{e:divtheo} on this term then leads to
\eqb{l}
\ds\int_{\sR} \Big(\rho\,\dot\phi + \phi\,\big(\dot\rho + \rho\,\dot J/J \big) - h_1 + j^\alpha_{;\alpha}\Big)\,\dif a = 0 \qquad \forall\,\sR\subset\sS \,.
\eqe
Defining $h^\ast_1 := h_1 - \phi\,h$ and making use of the localization theorem \eqref{e:localize} and ODE \eqref{e:ODE-rho}, then gives
\eqb{l}
\rho\,\dot\phi = h_1^\ast -j^\alpha_{;\alpha} \qquad \forall\,\bx\in\sS\,,
\label{e:PDE-c}\eqe
which is the governing ODE for the relative concentration $\phi$.
In order to determine $\phi(t)$, the initial condition $\phi = \phi_0$ at $t=0$ is required.
Interesting special cases for $h_1^\ast$ are $h^\ast_1 = h_1$ (the total mass is conserved), $h^\ast_1 = (1-\phi)\,h$ (only the mass of species 1 is increasing), $h^\ast_1 = -\phi\,h$ (only the mass of species 2 is increasing) and $h^\ast_1 = 0$ (the mass increase of species 1 and 2 has the ratio $\phi$ to $1-\phi$).

\subsection{Surface momentum balance}

Before exploring momentum balance, the stress and bending moments of the shell have to be introduced. 
As shown in Sec.~\ref{s:stress}, these can be expressed w.r.t.~the three configurations $\sS$, $\sS_0$ and $\hat\sS$ -- similar to the strains in Sec.~\ref{s:kin_s}.
Based on this, linear and angular momentum balance are then discussed in Secs.~\ref{s:lin} and \ref{s:ang}.

\subsubsection{Stress and moment tensors}\label{s:stress}

For shells, the Cauchy stress tensor takes the form
\eqb{l}
\bsig = N^{\alpha\beta}\,\ba_\alpha\otimes\ba_\beta + S^\alpha\,\ba_\alpha\otimes\bn\,,
\label{e:bsig}\eqe
with the in-plane membrane components $N^{\alpha\beta}$ and the out-of-plane shear components $S^\alpha$.
The traction vector on the boundary with normal $\bnu=\nu_\alpha\,\ba^\alpha$ then follows from Cauchy's formula
\eqb{l}
\bT = \bsig^\mrT \bnu\,.
\eqe
Introducing
\eqb{l}
\bT^\alpha := \bsig^\mrT \ba^\alpha\,,
\eqe
then leads to $\bT = \bT^\alpha \nu_\alpha$.
\\
For later reference the surface tension,
\eqb{l}
\gamma := \frac{1}{2}\tr_{\!\mrs}\,\bsig = \frac{1}{2}\,\bsig:\bone =  \frac{1}{2}N^{\alpha\beta} a_{\alpha\beta}\,,
\label{e:gamma}\eqe
and the deviatoric surface stress,
\eqb{l}
\bsig_\mathrm{dev} := \bsig - \gamma\,\bi\,,
\label{e:sig_dev}\eqe
are introduced.
The latter has the in-plane components
\eqb{l}
N_\mathrm{dev}^{\alpha\beta} := N^{\alpha\beta} - \gamma\,a^{\alpha\beta}
\label{e:Ndev}\eqe
in basis $\ba_\alpha$. 
The out-of-plane component $S^\alpha$ is identical for $\bsig_\mathrm{dev}$ and $\bsig$.
\\
The Cauchy stress describes the physical stress in configuration $\sS$.
It can be mapped to the (non-physical) second Piola-Kirchhoff stress tensor $\bS$ in configuration $\sS_0$ using the classical pull-back formula
\eqb{l}
\bS = J\tilde\bF^{-1}\bsig\tilde\bF^{-\mrT},
\label{e:S}\eqe
where $\tilde\bF := \bF + \bn\otimes\bN$ is the full 3D deformation gradient for Kirchhoff-Love kinematics.
In the same fashion we introduce the stress in configuration $\hat\sS$ by
\eqb{l}
\hat\bsig := J_\el\,\tilde\bF_{\!\el}^{-1}\bsig\tilde\bF_{\!\el}^{-\mrT}
\eqe
and note that
\eqb{l}
\bS = J_\inel\,\tilde\bF_{\!\inel}^{-1}\hat\bsig\tilde\bF_{\!\inel}^{-\mrT},
\eqe
where $\tilde\bF_{\!\el} := \bF_{\!\el} + \bn\otimes\hat\bn$ and $\tilde\bF_{\!\inel} := \bF_{\!\inel} + \hat\bn\otimes\bN$.
This lead to
\eqb{l}
\hat\bsig = \hat N^{\alpha\beta}\,\hat\ba_\alpha\otimes\hat\ba_\beta + \hat S^\alpha\,\hat\ba_\alpha\otimes\hat\bn
\label{e:hatbsig}\eqe
and
\eqb{l}
\bS = N_0^{\alpha\beta}\,\bA_\alpha\otimes\bA_\beta + S_0^\alpha\,\bA_\alpha\otimes\bN\,,
\eqe
where $N_0^{\alpha\beta}:=J_\inel\,\hat N^{\alpha\beta}$, $\hat N^{\alpha\beta}:=J_\el\,N^{\alpha\beta}$, $S^\alpha_0:=J_\inel\,\hat S^\alpha$ and $\hat S^\alpha:=J_\el\,S^\alpha$.
\\
Similar to the stress tensor $\bsig$ and the traction vector $\bT$, the bending moment tensor
\eqb{l}
\bmu = -M^{\alpha\beta}\,\ba_\alpha\otimes\ba_\beta
\eqe
and the moment vector
\eqb{l}
\bM = \bmu^\mrT \bnu = \bM^\alpha \nu_\alpha\,,
\eqe
with $\bM^\alpha = \bmu^\mrT \ba^\alpha$, are introduced in configuration $\sS$.
Just like $\bsig$, $\bmu$ can be pulled back to $\hat\sS$ and $\sS_0$ by
\eqb{l}
\hat\bmu := J_\el\,\tilde\bF_{\!\el}^{-1}\bmu\,\tilde\bF_{\!\el}^{-\mrT} = -\hat M^{\alpha\beta}\,\hat\ba_\alpha\otimes\hat\ba_\beta
\eqe
and
\eqb{l}
\bmu_0 := J\,\tilde\bF^{-1}\bmu\,\tilde\bF^{-\mrT} = -M_0^{\alpha\beta}\,\bA_\alpha\otimes\bA_\beta\,,
\eqe
where $M_0^{\alpha\beta}:=J_\inel\,\hat M^{\alpha\beta}$ and $\hat M^{\alpha\beta}:=J_\el\,M^{\alpha\beta}$.

\subsubsection{Linear surface momentum balance}\label{s:lin}

The linear surface momentum balance is given by
\eqb{l}
\ds\pad{}{t}\int_{\sR} \rho\,\bv\,\dif a = \int_\sR \bff\,\dif a + \int_{\partial\sR}\bT\,\dif s + \int_\sR h\,\bv\,\dif a \qquad \forall\,\sR\subset\sS\,,
\eqe
where $\bv:=\dot\bx$ is the current surface velocity, $\bff$ is a distributed surface load (force per current area) that, for two-species mixtures, has contributions acting on species 1 and 2 (i.e.~$\bff = \bff_{\!1} + \bff_{\!2}$), $\bT$ is the traction vector on boundary $\partial\sR$ and $h\,\bv$ accounts for the momentum change of the added mass. 
Applying Reynolds' transport theorem \eqref{e:Rey}, surface divergence theorem \eqref{e:divtheo}, localization \eqref{e:localize} and ODE \eqref{e:ODE-rho} gives
\eqb{l}
\bT^\alpha_{;\alpha} +\bff = \rho\,\dot\bv \qquad \forall\,\bx\in\sS\,,
\label{e:PDE-v}\eqe
which is the governing PDE for the motion. 
In order to determine $\bv(t)$, the initial condition $\bv = \bv_0$ at $t=0$ is required.
In order to determine $\bx(t)$, the additional initial condition $\bx = \bX$ at $t=0$ is required.
PDE \eqref{e:PDE-v} is exactly the same as for the mass conserving case, e.g.~see \citet{shelltheo}.

\textbf{Remark 3:} The surface load can also be defined per mass, i.e. $\bb := \bff/\rho$ and then decomposed as $\bff = \rho\,\bb = \rho_1\bb_1 + \rho_2\bb_2$ for the mixture. 
If $\bb_1 = \bb_2$, as for gravity, we find $\bb = \bb_1 = \bb_2$. 

\subsubsection{Angular surface momentum balance}\label{s:ang}

Also the angular surface momentum balance
\eqb{l}
\ds\pad{}{t}\int_{\sR} \bx\times\rho\,\bv\,\dif a = \int_\sR \bx\times\bff\,\dif a + \int_{\partial\sR}\big(\bx\times\bT+\bm\big)\,\dif s + \int_\sR \bx\times h\,\bv\,\dif a \qquad \forall\,\sR\subset\sS\,,
\eqe
where $\bm := \bn\times\bM$ is a distributed moment acting on boundary $\partial\sR$, leads to the same local equations as before, i.e.
\eqb{c}
S^\alpha = -M^{\beta\alpha}_{;\beta}\,, \\[1mm]
\sig^{\alpha\beta} := N^{\alpha\beta} - b^\beta_\gamma M^{\gamma\alpha} $ is symmetric$
\label{e:Sa_sigab}\eqe
for all $\bx\in\sS$, e.g.~see \citet{shelltheo}.

\subsection{Surface energy balance}

The surface energy balance can be expressed as
\eqb{l}
\!\ds\pad{}{t}\!\int_{\sR}\! \rho\,e\,\dif a = \int_\sR \!\bv\cdot\bff\,\dif a + \!\int_{\partial\sR}\!\big(\bv\cdot\bT+\dot\bn\cdot\bM\big)\,\dif s + \!\int_\sR\!h\,e\,\dif a + \!\int_\sR \!\rho\,r\,\dif a + \!\int_{\partial\sR}\!q_\nu\,\dif s \quad \forall\,\sR\subset\sS,
\label{e:bal-u}\eqe
where
\eqb{l}
e = u + \frac{1}{2}\bv\cdot\bv
\eqe
is the specific energy (per unit mass) at $\bx\in\sB$ that contains the stored energy $u$ and the kinetic energy $\bv\cdot\bv/2$.
The first two terms on the right hand side of \eqref{e:bal-u} account for the mechanical power of the external forces $\bff$ and $\bT$ and external moment $\bM$.
The third term accounts for the power required to add mass $h$: power is needed to bring the added mass to energy level $u$ and velocity $\bv$.\footnote{If the added mass carries initial, nonzero energy $e_0$, this energy contribution can be accounted for in the $\rho\,r$ term. Alternatively, if one does not wish to account for $e_0$ in $\rho\,r$, $h\,e$ in \eqref{e:bal-u} should be replaced by $h\,(e-e_0)$.} 
The last two terms account for the thermal power of an external heat source $r$ and a boundary influx $q_\nu$.
Defining the surface heat flux vector $\bq=q^\alpha\ba_\alpha$ through
\eqb{l}
q_\nu = -\bq\cdot\bnu = -q^\alpha\nu_\alpha\,,
\eqe
the surface divergence theorem gives
\eqb{l}
\ds\int_{\partial\sR}q_\nu\,\dif s = -\int_{\sR}q^\alpha_{;\alpha}\,\dif a\,.
\eqe
Using the surface divergence theorem on the $\bv\cdot\bT$ term gives
\eqb{l}
\ds\int_{\partial\sR}\bv\cdot\bT\,\dif s = \int_\sR\big(\bv\cdot\bT^\alpha_{;\alpha} + \frac{1}{2}\sigma^{\alpha\beta}\dot a_{\alpha\beta} + M^{\alpha\beta}\dot b_{\alpha\beta} \big)\,\dif a - \int_{\partial\sR}\dot\bn\cdot\bM\,\dif s\,,
\eqe
see \citet{sahu17}.
Using these two equations, ODE \eqref{e:ODE-rho} and PDE \eqref{e:PDE-v} then gives 
\eqb{l}
\rho\,\dot u =  \frac{1}{2}\sigma^{\alpha\beta}\dot a_{\alpha\beta} + M^{\alpha\beta}\dot b_{\alpha\beta} + \rho\,r - q^\alpha_{;\alpha}\,,
\qquad \forall\,\bx\in\sS\,,
\label{e:PDE-u}\eqe
which is the governing PDE for $u$. 
In order to determine $u(t)$, the initial condition $u = u_0$ at $t=0$ is required. 
PDE \eqref{e:PDE-u} has the same format as in the classical case when $h=0$ and no split of $\bF$ is considered \citep{sahu17}.
But due to the split of $\bF$, we can now write 
\eqb{l}
\rho\,\dot u =  \sigma^{\alpha\beta}\big(\dot\eps^\el_{\alpha\beta}+\dot\eps^\inel_{\alpha\beta}\big) + M^{\alpha\beta}\big(\dot\kappa^\el_{\alpha\beta}+\dot\kappa^\inel_{\alpha\beta}\big) + \rho\,r - q^\alpha_{;\alpha}\,,
\qquad \forall\,\bx\in\sS\,,
\label{e:PDE-u2}\eqe
according to eqs.~\eqref{e:deps}, \eqref{e:dkap}, \eqref{e:deps2} and \eqref{e:dkap2}.

\textbf{Remark 4:} The $\frac{1}{2}\sigma^{\alpha\beta}\dot a_{\alpha\beta}\,\dif a$ term can also be rewritten as
\eqb{l}
\frac{1}{2}\sigma^{\alpha\beta}\dot a_{\alpha\beta}\, \dif a = \bsig:\bD\,\dif a = \bS:\dot\bE\, \dif A\,,
\eqe
where
\eqb{lll}
\bD \is \frac{1}{2}\dot a_{\alpha\beta}\,\ba^\alpha \otimes \ba^\beta, \\[1mm]
\dot\bE \is \frac{1}{2}\dot a_{\alpha\beta}\,\bA^\alpha \otimes \bA^\beta,
\eqe
are the symmetric velocity gradient (e.g.~see \citet{cism}) and the Green-Lagrange strain rate (following from \eqref{e:E}), respectively, and $\bsig$ and $\bS$ are given by \eqref{e:bsig} and \eqref{e:S}. 
This illustrates that the stress component $\sig^{\alpha\beta}$ (and energy $\frac{1}{2}\sigma^{\alpha\beta}\dot a_{\alpha\beta}$) is expressed neither w.r.t. $\sS_0$ nor $\sS$, but directly w.r.t.~parameter space $\sP$ \citep{solidshell}.
$\bsig$ and $\bS$, on the other hand are specific to $\sS$ and $\sS_0$, respectively.

\textbf{Remark 5:} 
In \eqref{e:bal-u} the quantities $e$, $\bff$, $\bT$, $\bM$, $r$ and $q_\nu$ are defined for the common mixture in order to avoid dealing with partial quantities. 
In \citet{sahu17}, on the other hand, $\bff$ is defined partially, such that the second term in \eqref{e:bal-u} is the area integral over $\bv_1\cdot\rho_1\bb_1 + \bv_2\cdot\rho_2\bb_2$. 
This leads to an extra term in \eqref{e:PDE-u} and \eqref{e:PDE-u2} if $\bb_2\neq\bb_1$.

\subsection{Surface entropy balance}

The surface entropy balance is given by
\eqb{l}
\ds\pad{}{t}\int_{\sR} \rho\,s\,\dif a = \int_\sR\big(\rho\,\eta_\mre + \rho\,\eta_\mri + h\,s\big)\,\dif a + \int_{\partial\sR}\tilde q_\nu\,\dif s
\qquad \forall\,\sR\subset\sS\,,
\eqe
where $s$ is the specific entropy at $\bx\in\sS$, $\eta_\mre$ is the external entropy production rate caused by external loads and heat sources, $\tilde q_\nu$ is an entropy influx on the boundary of the surface, $h\,s$ accounts for the entropy increase due to the added mass,\footnote{If the added mass carries initial, nonzero entropy, this additional entropy contribution can be accounted for in the $\rho\,\eta_\mre$ term.} and $\eta_\mri$ is the internal entropy production rate, which according to the second law of thermodynamics satisfies
\eqb{l}
\eta_\mri\geq 0 \qquad \forall\,\bx\in\sS\,.
\label{e:2.law}\eqe
Defining the surface entropy flux vector $\tilde\bq=\tilde q^\alpha\ba_\alpha$ through
\eqb{l}
\tilde q_\nu = -\tilde\bq\cdot\bnu = -\tilde q^\alpha\nu_\alpha\,,
\eqe
the surface divergence theorem and the localization theorem lead to the local equation
\eqb{l}
\rho\,\dot s =  \rho\,\eta_e + \rho\,\eta_\mri - \tilde q^\alpha_{;\alpha}\,,
\qquad \forall\,\bx\in\sS\,,
\label{e:PDE-s}\eqe
which can be used to derive constitutive equations as is discussed in Sec.~\ref{s:consti}.
For this, we introduce the Helmholtz free energy $\psi:=u-Ts$, such that
\eqb{l}
\dot s = \big(\dot u - \dot Ts - \dot\psi\big)/T\,.
\eqe
Here $T>0$ is the absolute temperature.
Inserting this and PDE \eqref{e:PDE-u2} into \eqref{e:PDE-s} then gives
\eqb{l}
T\rho\,\dot s = \sigma^{\alpha\beta}\big(\dot\eps^\el_{\alpha\beta}+\dot\eps^\inel_{\alpha\beta}\big) + M^{\alpha\beta}\big(\dot\kappa^\el_{\alpha\beta}+\dot\kappa^\inel_{\alpha\beta}\big) + \ds \rho\,r - T\bigg(\frac{q^\alpha}{T}\bigg)_{\!\!;\alpha}\! - \frac{q^\alpha T_{;\alpha}}{T} - \rho\,\dot T s - \rho\,\dot\psi\,.
\label{e:2lawa}\eqe
In deriving this equation we have used local energy balance \eqref{e:PDE-u}, which in turn uses local mass balance \eqref{e:ODE-rho} and local momentum balance \eqref{e:PDE-v}.
We have thus used all PDEs apart from the local concentration balance \eqref{e:PDE-c}.
In order to account for it we add it to the right hand side of \eqref{e:2lawa} using the Lagrange multiplier method, i.e.
\eqb{l}
T\rho\,\dot s = ... + \mu\,\big(\rho\,\dot\phi-h_1^\ast+j^\alpha_{;\alpha}\big)\,,
\label{e:2lawb}\eqe
where $\mu$ is the Lagrange multiplier that corresponds to the chemical potential as will be shown later.
The last term in \eqref{e:2lawb} can be rewritten as
\eqb{l}
\mu\,j^\alpha_{;\alpha} = \ds T\bigg(\frac{\mu\,j^\alpha}{T}\bigg)_{\!\!;\alpha}\! - T\,j^\alpha\bigg(\frac{\mu}{T}\bigg)_{\!\!;\alpha}
\eqe
in order to combine it with the $T(q^\alpha/T)_{;\alpha}$ term.
We thus find
\eqb{lll}
T\rho\,\dot s \is \ds \rho\,r - \mu\,h^\ast_1 - T\bigg(\frac{q^\alpha}{T} - \frac{\mu\,j^\alpha}{T}\bigg)_{\!\!;\alpha}\! + 
\sigma^{\alpha\beta}\big(\dot\eps^\el_{\alpha\beta}+\dot\eps^\inel_{\alpha\beta}\big) 
+ M^{\alpha\beta}\big(\dot\kappa^\el_{\alpha\beta}+\dot\kappa^\inel_{\alpha\beta}\big) \\[4mm]
\mi \rho\,s\,\dot T - \ds\frac{q^\alpha T_{;\alpha}}{T} + \rho\,\mu\,\dot\phi - Tj^\alpha\bigg(\frac{\mu}{T}\bigg)_{\!\!;\alpha}\! - \rho\,\dot\psi\,.
\label{e:2lawc}\eqe
On the right hand side here, the only source terms are the first two terms, while the only divergence-like term is the third term.  
Comparing this with \eqref{e:PDE-s}, we can thus identify $\eta_\mre=r/T-\mu\,h^\ast_1/(T\rho)$ and $\tilde q^\alpha = q^\alpha/T-\mu\,j^\alpha/T$, such that the second law \eqref{e:2.law} yields
\eqb{l}
T\rho\,\eta_\mri = \ds \sigma^{\alpha\beta}\big(\dot\eps^\el_{\alpha\beta}+\dot\eps^\inel_{\alpha\beta}\big) + M^{\alpha\beta}\big(\dot\kappa^\el_{\alpha\beta}+\dot\kappa^\inel_{\alpha\beta}\big) - \rho\,s\,\dot T  - \frac{q^\alpha T_{;\alpha}}{T} + \rho\,\mu\,\dot\phi - Tj^\alpha\bigg(\frac{\mu}{T}\bigg)_{\!\!;\alpha}\! - \rho\,\dot\psi \geq 0 \,.
\label{e:2law}\eqe

\section{Problem statement}\label{s:prob}

In this work we consider the case of coupling elastic deformations with either growth, swelling, viscosity, plasticity or thermal deformation.
So there is only coupling of two deformation types.
In principle three and more types can also be coupled.
This would require introducing further intermediate configurations.
This section discusses the strong form for the coupled two-field problem. 
The recovery of the intermediate configuration is also addressed.

\subsection{Strong form}

The strong form can be unified by the problem statement: 
Find $\bx(\xi^\alpha,t)$, $\hat a_{\alpha\beta}(\xi^\gamma,t)$ and $\hat b_{\alpha\beta}(\xi^\gamma,t)$ satisfying PDE \eqref{e:PDE-v} and,
\begin{itemize}
\item\underline{for growth,} ODE \eqref{e:ODE-rho}. 
In this case, $\hat a_{\alpha\beta}$ and $\hat b_{\alpha\beta}$ are either prescribed or defined through~$\rho$. 
So the primary unknowns are $\bx$ and $\rho$.
Examples are given by Eqs.~\eqref{e:hata_dil}, \eqref{e:hatb_dil}, \eqref{ex:Jin1} and \eqref{ex:kin1}.
\item\underline{for swelling,} PDE \eqref{e:PDE-c}. 
In this case, $\hat a_{\alpha\beta}$ and $\hat b_{\alpha\beta}$ are  defined through $\phi$, e.g.~by \eqref{e:hata_dil}, \eqref{e:hatb_dil}, \eqref{ex:Jin2} and \eqref{ex:kin2}, and so the primary unknowns are $\bx$ and $\phi$. If the swelling is not mass conserving (i.e.~$h\neq0)$, $\rho$ is also unknown and needs to be obtained from ODE \eqref{e:ODE-rho}. 
\item\underline{for viscoelasticity and elastoplasticity,} an evolution law (ODE) for $\hat a_{\alpha\beta}$ and $\hat b_{\alpha\beta}$, like \eqref{e:evol1}, \eqref{e:evol4}, \eqref{e:evol5} or \eqref{e:evol7}.
\item\underline{for thermoelasticity,} PDE \eqref{e:PDE-u2}. 
In this case, $\hat a_{\alpha\beta}$ and $\hat b_{\alpha\beta}$ are defined through $T$, e.g.~by \eqref{e:hata_dil}, \eqref{e:hatb_dil}, \eqref{ex:Jin3} and \eqref{ex:kin4}, and so the primary unknowns are $\bx$ and $T$.
\end{itemize}
In general, the governing ODEs and PDEs are nonlinear and coupled, and hence need to be solved numerically.
The ODEs can be solved locally using numerical integration schemes like the implicit Euler scheme.
The problem simplifies if $\rho$, $\phi$ or $T$ are prescribed.
In those cases the problem decouples.
In order to fully characterize PDEs \eqref{e:PDE-c}, \eqref{e:PDE-v} and \eqref{e:PDE-u}, constitutive expressions for the mass flux, stress, bending moments and heat flux are needed. 
Those are discussed in Sec.~\ref{s:consti}.

\subsection{Recovery of $\hat\ba_\alpha$}

Strictly the recovery of $\hat\ba_1$ and $\hat\ba_2$, which fully define the intermediate configuration $\hat\sS$, is not needed to solve the problem, but it may still be interesting to reconstruct $\hat\ba_\alpha$, and from it $\bF_{\!\inel}$, for various reasons.
\\
The recovery is straightforward for isotropic growth, isotropic swelling and isotropic thermal expansion, since in these cases
$\hat\ba_\alpha$ is given by \eqref{e:hatba_dil},
with $\lambda_\inel$ being either prescribed directly or defined through $\rho,$ $\phi$ or $T$, as in some of the examples of Sec.~\ref{s:ex}.
\\
For viscoelasticity and elastoplasticity, on the other hand, the two vectors $\hat\ba_1$ and $\hat\ba_2$ can be determined from the two equations
\eqb{lll}
\hat a_{\alpha\beta} \is \hat\ba_\alpha\cdot\hat\ba_\beta \\[1mm]
\hat b_{\alpha\beta} \is \hat\ba_{\alpha,\beta}\cdot\hat\bn
\eqe
that each have three cases. 
In order to eliminate rigid body rotations, $\hat\ba_1$ and $\hat\ba_2$ need to be fixed at some point.

\textbf{Remark 6}: If no inelastic bending occurs, i.e. $\hat\bn = \bN$, the second equation can be replaced by the scalar equation
\eqb{l}
\hat\ba_\alpha\cdot\bN = 0\,,
\eqe
that has two cases and fixes the inclination of $\hat\sS$, and the condition
\eqb{l}
(\hat\ba_1\times\hat\ba_2)\cdot\bN = \norm{\hat\ba_1\times\hat\ba_2}
\eqe
that fixes the orientation of $\hat\sS$.
Additionally, $\hat\ba_1$ (or $\hat\ba_2$) needs to be fixed at a point to eliminate the rigid body rotation around $\bN$.

\section{Constitution}\label{s:consti}

This section derives the constitutive equations following from the second law of thermodynamics and provides various examples for growth, swelling, elasticity, viscosity, plasticity and thermal expansion.

\subsection{Constitutive theory}

In general, the Helmholtz free energy is supposed to be a function of the elastic strains $\eps^\el_{\alpha\beta}$ and $\kappa^\el_{\alpha\beta}$, temperature $T$ and concentration $\phi$, i.e. 
\eqb{l}
\psi = \psi\big(\eps^\el_{\alpha\beta},\kappa^\el_{\alpha\beta},T,\phi\big)\,,
\eqe
such that
\eqb{l}
\dot\psi = \ds\pa{\psi}{\eps^\el_{\alpha\beta}}\,\dot\eps^\el_{\alpha\beta} + \pa{\psi}{\kappa^\el_{\alpha\beta}}\,\dot\kappa^\el_{\alpha\beta} + \pa{\psi}{T}\,\dot T
+ \pa{\psi}{\phi}\,\dot\phi \,.
\eqe
Eq.~\eqref{e:2law} then yields\footnote{Replacing $\dot\eps^\el_{\alpha\beta}$ according to (\ref{e:deps2}.1), Eq.~\eqref{e:2law2} can be also expressed in terms of $\dot\eps_{\alpha\beta}$ and $\dot\eps^\inel_{\alpha\beta}$.}
\eqb{lllll}
T\rho\,\eta_\mri \is \ds \bigg(\sig^{\alpha\beta}-\rho\pa{\psi}{\eps^\el_{\alpha\beta}}\bigg)\dot\eps^\el_{\alpha\beta} \plus \sig^{\alpha\beta}\dot\eps^\inel_{\alpha\beta} \\[4.5mm]
\plus \ds\bigg(M^{\alpha\beta}-\rho\pa{\psi}{\kappa^\el_{\alpha\beta}}\bigg)\dot\kappa^\el_{\alpha\beta} \plus M^{\alpha\beta}\dot\kappa^\inel_{\alpha\beta} \\[4.5mm]
\mi \ds\rho\bigg(s+\pa{\psi}{T}\bigg)\dot T \mi T\ds\frac{q^\alpha T_{;\alpha}}{T} \\[4.5mm]
\plus \ds\rho\bigg(\mu-\pa{\psi}{\phi}\bigg)\dot\phi \mi T\,j^\alpha\Big(\ds\frac{\mu}{T}\Big)_{;\alpha}  ~ \geq 0 \,.
\label{e:2law2}\eqe
We now invoke the procedure of \citet{coleman64}.
Two cases have to be considered.
The first case supposes that $\eps^\inel_{\alpha\beta}$ and $\kappa^\inel_{\alpha\beta}$ are independent process variables. 
Then, since \eqref{e:2law2} is true for all rates $\dot\eps^\el_{\alpha\beta}$, $\dot\eps^\inel_{\alpha\beta}$, $\dot\kappa^\el_{\alpha\beta}$, $\dot\kappa^\inel_{\alpha\beta}$ $\dot T$, $\dot\phi$ and gradients $T_{;\alpha}$, $(\mu/T)_{;\alpha}$, we obtain the sufficient conditions\footnote{They are not necessary conditions as they can be combined into new conditions.}
\eqb{rllrll}
\sig^{\alpha\beta} \is \ds\rho\pa{\psi}{\eps^\el_{\alpha\beta}}\,, & \sig^{\alpha\beta}\dot\eps^\inel_{\alpha\beta} \isgeq 0\,, \\[4.5mm]
M^{\alpha\beta} \is \ds\rho\pa{\psi}{\kappa^\el_{\alpha\beta}}\,, & M^{\alpha\beta}\dot\kappa^\inel_{\alpha\beta} \isgeq 0\,, \\[4.5mm]
s\is -\ds\pa{\psi}{T}\,, & q^\alpha\,T_{;\alpha} \isleq 0\,, \\[3.5mm]
\mu \is \ds\pa{\psi}{\phi}\,, & j^\alpha \bigg(\ds\frac{\mu}{T}\bigg)_{\!\!;\alpha} \isleq 0\,,
\label{e:consti1a}\eqe
which are the general constitutive relations for the stresses $\sig^{\alpha\beta}$, bending moments $M^{\alpha\beta}$, inelastic strains $\eps^\inel_{\alpha\beta}$, inelastic curvature change $\kappa^\inel_{\alpha\beta}$, heat flux $q^\alpha$, entropy $s$, concentration flux $j^\alpha$ and Lagrange multiplier $\mu$.
The latter is identified to be the chemical potential.
\\
On the other hand, if $\dot\eps^\inel_{\alpha\beta}$ and $\dot\kappa^\inel_{\alpha\beta}$ are functions of $T$ and $\phi$, their rates can be expanded into
\eqb{lll}
\dot\eps^\inel_{\alpha\beta} \is \ds\pa{\eps^{\inel}_{\alpha\beta}}{T}\,\dot T + \pa{\eps^{\inel}_{\alpha\beta}}{\phi}\,\dot\phi\,, \\[4.5mm]
\dot\kappa^\inel_{\alpha\beta} \is \ds\pa{\kappa^{\inel}_{\alpha\beta}}{T}\,\dot T + \pa{\kappa^{\inel}_{\alpha\beta}}{\phi}\,\dot\phi\,.
\eqe
Inserting this into \eqref{e:2law2} then yields
\eqb{lll}
T\rho\,\eta_\mri 
\is \ds \bigg(\sig^{\alpha\beta}-\rho\pa{\psi}{\eps^\el_{\alpha\beta}}\bigg)\dot\eps^\el_{\alpha\beta} 
+ \ds\bigg(M^{\alpha\beta}-\rho\pa{\psi}{\kappa^\el_{\alpha\beta}}\bigg)\dot\kappa^\el_{\alpha\beta} \\[4.5mm]
\mi \ds\rho\bigg(s+\pa{\psi}{T} - \frac{\sig^{\alpha\beta}}{\rho}\pa{\eps^{\inel}_{\alpha\beta}}{T} - \frac{M^{\alpha\beta}}{\rho}\pa{\kappa^{\inel}_{\alpha\beta}}{T}\bigg)\dot T ~-~ T\ds\frac{q^\alpha T_{;\alpha}}{T} \\[4.5mm]
\plus \ds\rho\bigg(\mu-\pa{\psi}{\phi} + \frac{\sig^{\alpha\beta}}{\rho}\pa{\eps^{\inel}_{\alpha\beta}}{\phi} + \frac{M^{\alpha\beta}}{\rho}\pa{\kappa^{\inel}_{\alpha\beta}}{\phi}\bigg)\dot\phi ~-~ T\,j^\alpha\Big(\ds\frac{\mu}{T}\Big)_{;\alpha}  ~ \geq 0 \,.
\label{e:2law3}\eqe
Since this is true for all $\dot\eps^\el_{\alpha\beta}$, $\dot\kappa^\el_{\alpha\beta}$, $\dot T$, $\dot\phi$, $T_{;\alpha}$ and $(\mu/T)_{;\alpha}$, we now find
\eqb{rll}
s\is -\ds\pa{\psi}{T} + \frac{\sig^{\alpha\beta}}{\rho}\pa{\eps^{\inel}_{\alpha\beta}}{T} + \frac{M^{\alpha\beta}}{\rho}\pa{\kappa^{\inel}_{\alpha\beta}}{T} \,, \\[3.5mm]
\mu \is \ds\pa{\psi}{\phi} - \frac{\sig^{\alpha\beta}}{\rho}\pa{\eps^{\inel}_{\alpha\beta}}{\phi} - \frac{M^{\alpha\beta}}{\rho}\pa{\kappa^{\inel}_{\alpha\beta}}{\phi}\,,
\label{e:consti1b}\eqe
together with the equations for $\sig^{\alpha\beta}$ and $M^{\alpha\beta}$ and the inequality conditions for $q^\alpha$ and $j^\alpha$ already listed in \eqref{e:consti1a}.
Now, the conditions $\sig^{\alpha\beta}\dot\eps^\inel_{\alpha\beta} \geq 0$ and $M^{\alpha\beta}\dot\kappa^\inel_{\alpha\beta} \geq 0$ are no longer a requirement.
As \eqref{e:consti1b} shows for the second case, the entropy and chemical potential have contributions coming from the inelastic deformation measures $\eps^{\inel}_{\alpha\beta}$ and $\kappa^{\inel}_{\alpha\beta}$.

\subsection{Alternative constitutive description}

By redefining the Helmholtz free energy, we can rewrite some of the above constitutive equations.
Since the Helmholtz free energy $\psi$ is defined per unit mass, the total energy is
\eqb{l}
\Pi := \ds\int_\sM \psi\,\dif m = \int_\sS\rho\,\psi\,\dif a\,,
\label{e:Pi}\eqe
where the first integral denotes the integration over the total mass of surface $\sS$.
Defining $\rho^0$ as the current density in the reference configuration, i.e.~$\rho^0:=J\,\rho$, we can also write
\eqb{l}
\Pi = \ds\int_{\sS_0}\rho^0\,\psi\,\dif A\,.
\eqe
Due to growth ($h\neq0$), density $\rho^0$ is changing over time and is not equal to the initial density $\rho_0$ (unless $h=0$). 
Likewise, we introduce $\hat\rho$ as the density in $\hat\sS$, i.e.~$\hat\rho := J_\el\,\rho$, so that we can further write
\eqb{l}
\Pi = \ds\int_{\hat\sS}\hat\rho\,\psi\,\dif\hat a =  \int_{\hat\sS}\hat\Psi\,\dif\hat a\,,
\eqe
where $\dif\hat a = J_\inel\,\dif A$ and 
\eqb{l}
\hat\Psi:=\hat\rho\,\psi
\label{e:Psihat}\eqe
is the Helmholtz free energy per unit intermediate area. 
Since $\hat\rho$ is independent of the elastic deformation (as long as $\hat h=J_\el\,h$ is),\footnote{We can rewrite ODE \eqref{e:ODE-rho} into $\dot{\hat\rho} + \hat\rho\,\dot J_\inel/J_\inel - \hat h = 0$.} we can rewrite the constitutive laws for $\sig^{\alpha\beta}$ and $M^{\alpha\beta}$ into
\eqb{lll}
\sig^{\alpha\beta}_{(\el)} \is \ds\frac{1}{J_\el}\pa{\hat\Psi}{\eps^\el_{\alpha\beta}}\,, \\[4.5mm]
M^{\alpha\beta}_{(\el)} \is \ds\frac{1}{J_\el}\pa{\hat\Psi}{\kappa^\el_{\alpha\beta}}\,.
\label{e:consti2}\eqe
Subscript ``(el)" is added here to indicate that this is the stress following from the elasticity model. 
But since $\sig^{\alpha\beta}_{(\el)} = \sig^{\alpha\beta}_{(\inel)} = \sig^{\alpha\beta}$, brackets on this subscript are used.  
Using the alternative stress measures introduced in \eqref{e:hatbsig} and using identities \eqref{d:eps2} and \eqref{d:kap2}, we can further write
\eqb{lllll}
\hat\sig^{\alpha\beta}_{(\el)} \is \ds\pa{\hat\Psi}{\eps^\el_{\alpha\beta}} \is \ds\pa{\hat\Psi}{\eps_{\alpha\beta}} \,, \\[4.5mm]
\hat M^{\alpha\beta}_{(\el)} \is \ds\pa{\hat\Psi}{\kappa^\el_{\alpha\beta}} \is \ds\pa{\hat\Psi}{\kappa_{\alpha\beta}}\,.
\label{e:consti3}\eqe

\subsection{Constitutive examples}\label{s:ex}

The following subsections give examples for the elastic and inelastic material behavior of curved surfaces resulting from the constitutive laws in \eqref{e:consti1a}, \eqref{e:consti1b}  and \eqref{e:consti2}.
We therefore consider that the Helmholtz free energy has additive mechanical, thermal and concentrational parts, i.e.
\eqb{l}
\psi = \psi_\mathrm{mech} + \psi_\mathrm{therm} + \psi_\mathrm{conc}\,.
\eqe
An additional energy due to growth is not required since the energy change due to mass changes is already accounted for in $\Pi$ through $\rho$, see \eqref{e:Pi}.

\subsubsection{Growth models}\label{s:growth}

\textbf{i. Isotropic in-plane growth:}
In this case $\hat\ba_\alpha$ and $\bF_{\!\inel}$ are given by \eqref{e:hatba_dil} and \eqref{e:Fin_dil}.
If this growth is unrestricted and maintains constant density over time, i.e.~$J_\el=1$ and $\rho = \rho_0 =$ const. $\forall\,t$, ODE \eqref{e:ODE-rho} leads to the exponential growth law
\eqb{l}
J_\inel = J_0\,\exp(h\,t/\rho_0)\,,
\label{ex:Jin1}\eqe
where $J_0$ is a dimensionless constant. 
In case of restricted growth at changing density ($J_\el\neq 1$, $\rho\neq$ const.), expression \eqref{ex:Jin1} can still be used as a possible model.
However in that case, also other growth models in the form
\eqb{l}
J_\inel = J_\inel(t)
\eqe
are possible.
Examples are linear\footnote{w.r.t.~stretch $\lambda_\inel = \sqrt{J_\inel}$} growth in $h$,
\eqb{l}
J_\inel = J_0\,\big(1+c\,h\,t\big)^2\,,
\eqe
and logarithmic expansion in time,
\eqb{l}
J_\inel = J_0\,\big(1+\ln(1+t/t_0)\big)\,,
\label{ex:Jin1a}\eqe
where $J_0$, $c$ and $t_0$ are constants.
No matter what growth model is used, if $\rho$ is not assumed constant, it has to be solved for from ODE \eqref{e:ODE-rho}.
If growth is mass conserving, e.g.~during expansion \eqref{ex:Jin1a}, $\rho = \rho_0/J$ solves ODE \eqref{e:ODE-rho}.
Given $J_\inel$, $\hat a_{\alpha\beta}$ is then fully defined via \eqref{e:hata_dil}.

\textbf{ii. Curvature growth:}
An example for curvature growth is the isotopic bending model \eqref{e:hatk_dil} with the linear increase
\eqb{l}
\bar\kappa_\inel = 1+c\,h\,t\,,
\label{ex:kin1}\eqe
that could be caused by a one-sided mass source $h$.
$\hat b_{\alpha\beta}$ is then given by \eqref{e:hatb_dil}.

\subsubsection{Models for concentration induced swelling and diffusion}\label{s:conz}

\textbf{i. Linear isotropic swelling:} 
A classical model for swelling is the linear model
\eqb{l}
J_\inel = \lambda_\inel^2\,,\quad \lambda_\inel = 1 + \alpha_\mrc\,(\phi-\phi_0)\,,
\label{ex:Jin2}\eqe
where the material constant $\alpha_\mrc$ denotes the coefficient of chemical swelling.
Without loss of generality, one can then use $\hat a_{\alpha\beta} = J_\inel\,A_{\alpha\beta}$ as discussed in Sec.~\ref{s:indil}.

\textbf{ii. Chemical bending:}
An example for concentration induced curvature increase is the isotopic bending model \eqref{e:hatk_dil} with the linear curvature increase
\eqb{l}
\bar\kappa_\inel = 1 + \alpha_\kappa\,(\phi-\phi_0) \,,
\label{ex:kin2}\eqe
where $\alpha_\kappa$ is a constant.
This curvature increase could be caused by a one-sided swelling, e.g.~due to the binding of molecules to one side of a flexible membrane \citep{sahu17}.\\
Another, less trivial, example is a curvature increase due to a concentration difference between top and bottom surface, i.e.
\eqb{l}
\bar\kappa_\inel = 1 + \alpha_\kappa\,\big(\phi_{+}-\phi_{-}\big)\,,
\label{ex:kin3}\eqe
where $\phi_{+}$ and $\phi_{-}$ denote the top and bottom concentrations of surface $\sS$, respectively.
These need to be defined in a suitable way, e.g.~by using two separate PDEs of type \eqref{e:PDE-c} for the top and bottom surface.

\textbf{iii. Surface mass diffusion:}
A simple surface diffusion model satisfying \eqref{e:consti1a} is 
\eqb{l}
j^\alpha = -M\,a^{\alpha\beta}\,\bigg(\ds\frac{\mu}{T}\bigg)_{\!\!;\beta}\!\,,
\eqe
where $M$ is a constant.
Choosing
\eqb{l}
\psi = \ds\frac{c_\phi}{2}T\,\phi^2\,,
\label{e:psiphi}\eqe
where $c_\phi$ is a constant, we find the chemical potential
\eqb{l}
\mu = c_\phi\,T\,\phi - \ds\frac{1}{\rho\,J}\Big(2\alpha_\mrc\,\gamma_0\,\lambda_\inel + 4\alpha_\mrc\,\gamma_0^\mathrm{M}\,\lambda_\inel\,\bar\kappa_\inel + 2\alpha_{\kappa}\,\gamma_0^\mathrm{M}\,\lambda^2_\inel \Big)\,,
\label{e:full-mu}\eqe
due to \eqref{e:consti1b}, \eqref{e:hata_dil}, \eqref{e:hatb_dil}, \eqref{ex:Jin2} and \eqref{ex:kin2}.
Here $\gamma_0:=\sig^{\alpha\beta}_0 A_{\alpha\beta}/2$ and $\gamma_0^\mathrm{M}:=M^{\alpha\beta}_0 B_{\alpha\beta}/2$ follow from the stress definitions in Eq.~\eqref{e:Sa_sigab} and Sec.~\ref{s:stress}.
For the special case that $\gamma_0$, $\gamma_0^\mathrm{M}$ and $T$ are constant across $\sS$, we arrive at Fick's law
\eqb{l}
j^\alpha = -D\, a^{\alpha\beta}\,\phi_{;\beta}\,,
\label{e:Fick}\eqe
where $D = \tilde c_\phi\,M$, with $\tilde c_\phi = c_\phi - \big(2\alpha_\mrc^2\,\gamma_0 + 4\alpha_\mrc^2\,\gamma_0^\mathrm{M}\,\bar\kappa_\inel + 8\alpha_\mrc\,\alpha_\kappa\,\gamma_0^\mathrm{M}\,\lambda_\inel\big)/(T\,\rho\,J)$,
is the surface diffusivity.

\subsubsection{Mechanical membrane models}\label{s:exM}

This section discusses mechanical material models for elastic, viscous and plastic membrane behavior, for which bending moments are neglected. 
In this case we have $N^{\alpha\beta} = \sig^{\alpha\beta}$ according to \eqref{e:Sa_sigab}.

\textbf{i. Surface elasticity:} 
An example for the elastic response is the potential
\eqb{l}
\hat\Psi = \ds\frac{\Lambda}{4}\big(J_\el^2-1-2\ln J_\el\big) + \frac{G}{2}\big(I_1^\el - 2 - 2\ln J_\el\big)\,,
\label{e:Psi1}\eqe
which is adapted from the classical 3D Neo-Hookean material model \citep{shelltheo}.
The parameters $\Lambda$ and $G$ are material constants. 
From \eqref{e:Psi1} follows the membrane stress
\eqb{l}
\sig^{\alpha\beta}_{(\el)} = \ds\frac{\Lambda}{2J_\el}\big(J_\el^2-1\big)\,a^{\alpha\beta} + \frac{G}{J_\el}\big(\hat a^{\alpha\beta} - a^{\alpha\beta}\big)\,,
\label{e:sige1}\eqe
according to \eqref{e:consti2}, \eqref{d:Jel_Eel} and \eqref{d:Iel_Eel}.
The two terms in \eqref{e:Psi1} do not properly split dilatational and deviatoric energies.  
Such a split is achieved by the alternative model
\eqb{l}
\hat\Psi = \ds\frac{K}{4}\big(J_\el^2-1-2\ln J_\el\big) + \frac{G}{2}\bigg(\frac{I_1^\el}{J_\el}-2\bigg)\,,
\label{e:Psi2}\eqe
which is adapted from \citet{shelltheo}. 
The constants $K$ and $G$ denote the in-plane bulk and shear moduli. 
From \eqref{e:consti2} now follows the membrane stress
\eqb{l}
\sig^{\alpha\beta}_{(\el)} = \ds\frac{K}{2J_\el}\big(J_\el^2-1\big)\,a^{\alpha\beta} + \frac{G}{J^2_\el}\bigg(\hat a^{\alpha\beta} - \frac{I_1^\el}{2}a^{\alpha\beta}\bigg) \,.
\label{e:sige2}\eqe
Here the first part is purely dilatational, while the second is purely deviatoric.
Hence, the surface tension only depends on the first part, while the deviatoric stress only depends on the second part: 
From \eqref{e:gamma} and \eqref{e:sig_dev} follow the surface tension
\eqb{l}
\gamma = \ds\frac{K}{2J_\el}\big(J_\el^2-1\big)
\eqe
and the deviatoric stress
\eqb{l}
\sig_\mathrm{dev}^{\alpha\beta} = \ds\frac{G}{J^2_\el}\bigg(\hat a^{\alpha\beta} - \frac{I_1^\el}{2}a^{\alpha\beta}\bigg) \,.
\eqe
A third elasticity example is the linear elastic membrane model 
\eqb{l}
\hat\Psi = \ds\frac{1}{2}\eps^\el_{\alpha\beta}\,\hat c^{\alpha\beta\gamma\delta}\,\eps^\el_{\gamma\delta}\,,
\label{e:Psi3}\eqe
with the material tangent
\eqb{l}
\hat c^{\alpha\beta\gamma\delta} := \Lambda\, \hat a^{\alpha\beta} \hat a^{\gamma\delta} + G\,\big(\hat a^{\alpha\gamma}\hat a^{\beta\delta}+\hat a^{\alpha\delta}\hat a^{\beta\gamma}\big)
\label{e:cabcd}\eqe
based on the constants $\Lambda$ and $G$.
\eqref{e:Psi3} is analogous to the 3D St.-Venant-Kirchhoff material, from which it can be derived
\footnote{A 3D energy density needs to be multiplied by the shell thickness in order to obtain the membrane energy density $\hat\Psi$.}.
From \eqref{e:consti3} now follows
\eqb{l}
\hat\sig^{\alpha\beta}_{(\el)} = \hat c^{\alpha\beta\gamma\delta}\,\eps^\el_{\gamma\delta}\,,
\label{e:sige3}\eqe
which can be expanded into
\eqb{l}
\hat\sig^{\alpha\beta}_{(\el)} = \ds\frac{\Lambda}{2}\big(I_1^\el-2\big)\hat a^{\alpha\beta} + G\big(\hat a^{\alpha\gamma}\,a_{\gamma\delta}\,\hat a^{\beta\delta}-\hat a^{\alpha\beta} \big)\,.
\eqe

\textbf{ii. Surface viscosity:}
A simple shear viscosity model satisfying \eqref{e:consti1a} is
\eqb{l}
\hat\sig^{\alpha\beta}_{(\inel)} = -\eta\,\dot{\hat a}^{\alpha\beta}\,,
\label{e:sigi1}\eqe
where the material constant $\eta\geq0$ denotes the in-plane shear viscosity.\footnote{Proof: Using $\dot{\hat a}^{\alpha\beta} = -\hat a^{\alpha\gamma}\dot{\hat a}_{\gamma\delta}\,\hat a^{\beta\delta}$ and $2\hat\bD:=\dot{\hat a}_{\alpha\beta}\,\hat\ba^\alpha\otimes\hat\ba^\beta$ gives $\sig^{\alpha\beta}_{(\inel)}\dot\eps^\inel_{\alpha\beta} = 2\eta\,\hat\bD:\hat\bD/J_\el \geq 0$ for $\eta\geq0$.}
It is noted, that model \eqref{e:sigi1} is not purely deviatoric, since it generally leads to non-zero surface tension ($\gamma = \frac{1}{2}\,\sig^{\alpha\beta}\,a_{\alpha\beta}\neq 0$).
Another simple viscosity model satisfying \eqref{e:consti1a} is
\eqb{l}
\hat\sig^{\alpha\beta}_{(\inel)} = \lambda \,\dot J_\inel\,\hat a^{\alpha\beta}\,,
\label{e:sigi2}\eqe
where the material constant $\lambda\geq0$ denotes the in-plane bulk viscosity.\footnote{Proof: From \eqref{e:dJin2} follows 
$\sig^{\alpha\beta}_{(\inel)}\dot\eps^\inel_{\alpha\beta} = 2\lambda\,\dot J^2_\inel/J\geq 0$ for $\lambda\geq0$.} It is noted, that model \eqref{e:sigi2} is not purely dilatational, since it generally leads to non-zero shear stresses.
\\
If there is no elastic deformation, $\hat a_{\alpha\beta} = a_{\alpha\beta}$.
If there is elastic deformation, $\hat a_{\alpha\beta}$ has to be determined from an evolution law.
This depends on the rheological model considered.
\begin{figure}[h]
\begin{center} \unitlength1cm
\begin{picture}(0,2.3)
\put(-7.0,0.0){\includegraphics[width=40mm]{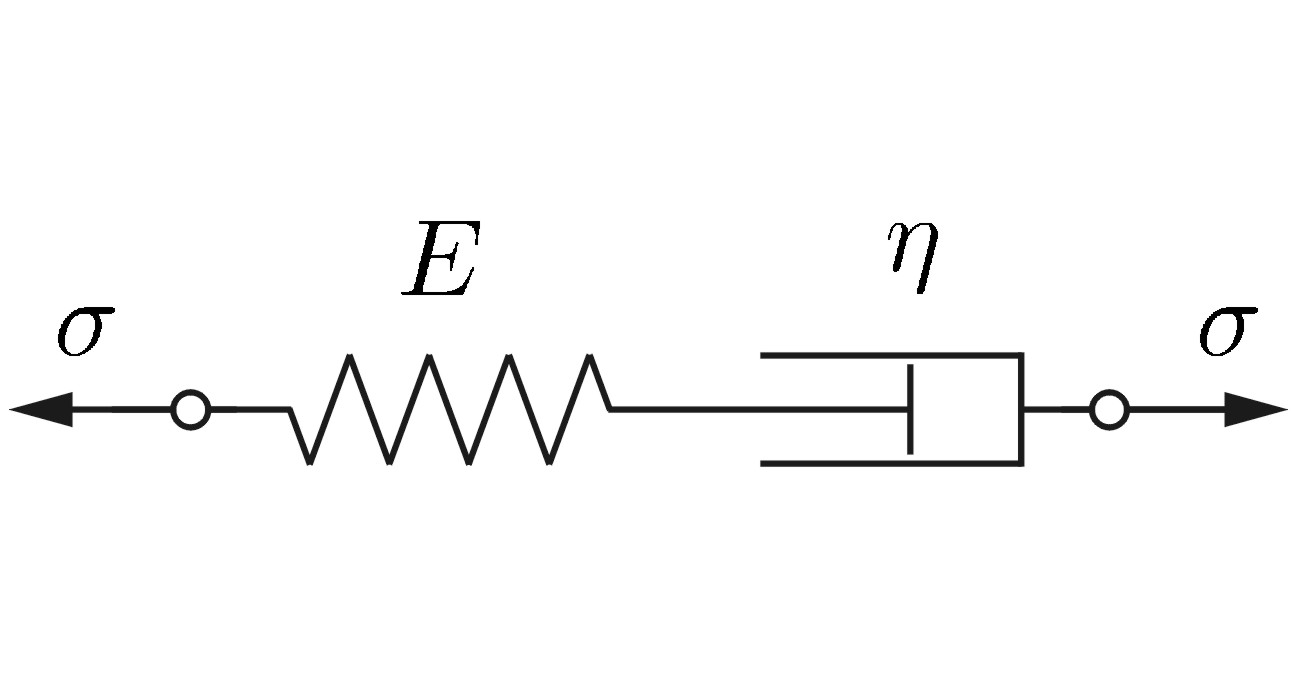}}
\put(-1.7,-.25){\includegraphics[width=40mm]{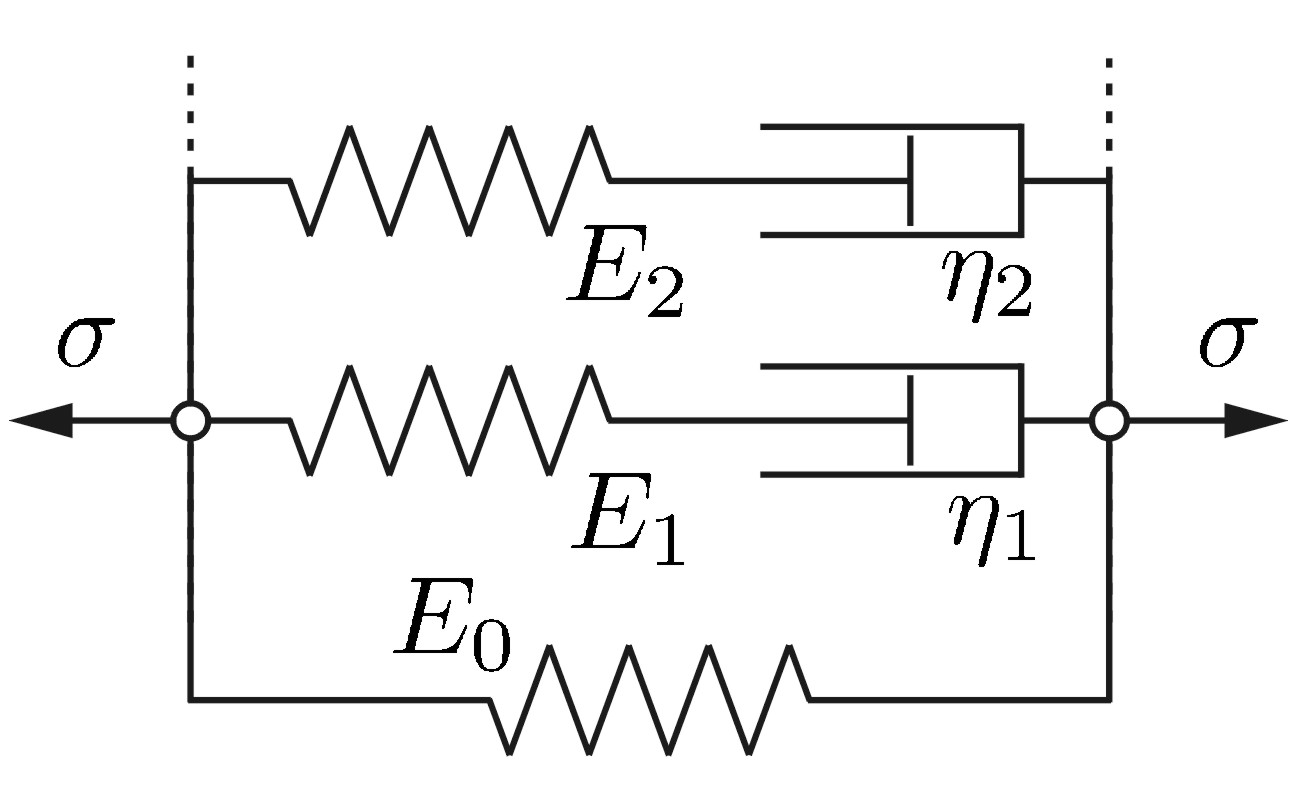}}
\put(3.2,0.0){\includegraphics[width=40mm]{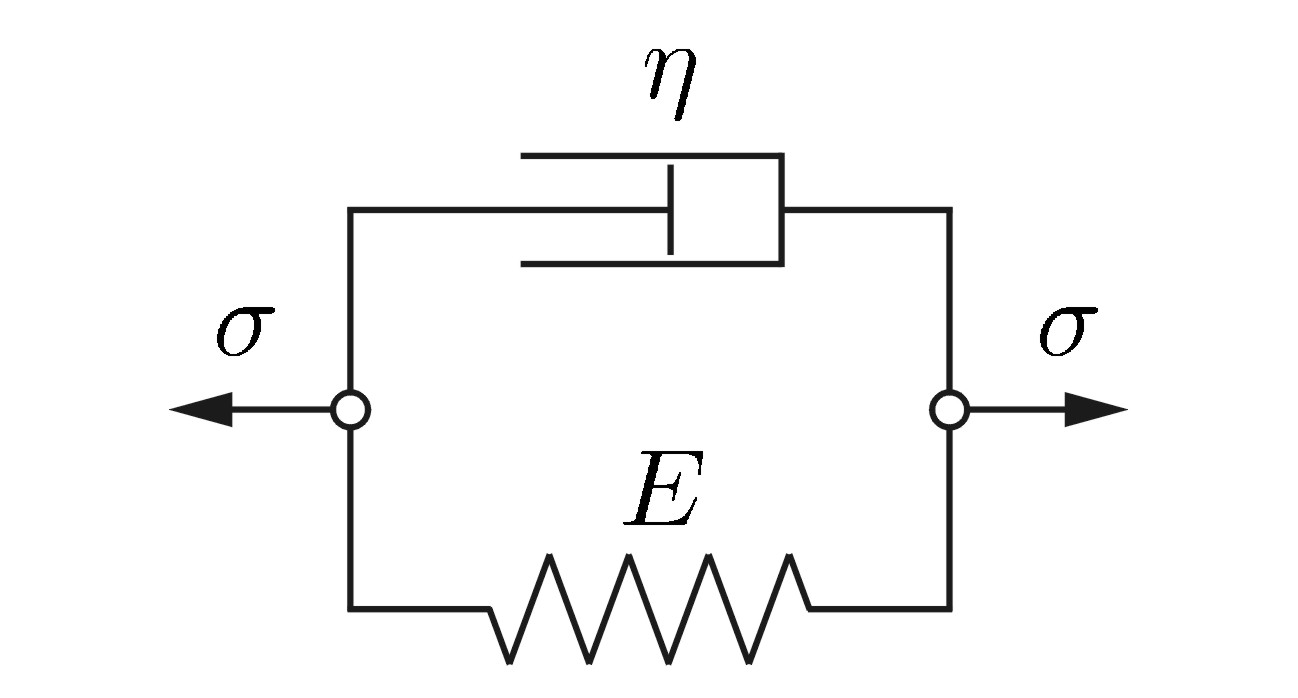}}
\put(-7.1,0){a.}
\put(-1.85,0){b.}
\put(3.65,0){c.}
\end{picture}
\caption{Surface rheology: a.~viscoelastic Maxwell fluid; b.~generalized viscoelastic solid; c.~viscoelastic Kelvin solid. Even though these are 1D models, they can be used in 2D and 3D, e.g.~by applying them to dilatation and shear. $E$ then plays the role of bulk and shear modulus.}
\label{f:rheo}
\end{center}
\end{figure}
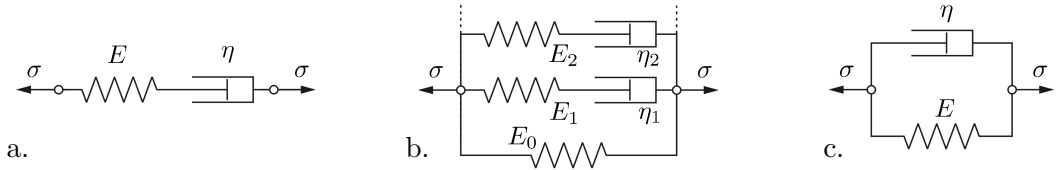
For a Maxwell element, see Fig.~\ref{f:rheo}a, the evolution law for $\hat a^{\alpha\beta}$ follows from
\eqb{l}
\sig_{(\el)}^{\alpha\beta}\big(\hat a^{\gamma\delta}\big) = \sig_{(\inel)}^{\alpha\beta}\big(\hat a^{\gamma\delta},\dot{\hat a}^{\gamma\delta}\big)\,,
\label{e:evol0}\eqe
which is a nonlinear ODE that can be solved locally using numerical methods (e.g.~implicit Euler).
For example, for models \eqref{e:sige1} and \eqref{e:sigi1}, the evolution law is 
\eqb{l}
\dot{\hat a}^{\alpha\beta} 
= \ds\frac{\Lambda}{2\eta}\big(1-J_\el^2\big)\,a^{\alpha\beta} + \frac{G}{\eta}\Big(a^{\alpha\beta}-\hat a^{\alpha\beta}\Big) \,,
\label{e:evol1}\eqe
where $J_\el$ is a function of $\hat a^{\alpha\beta}$ according to \eqref{e:Jei}.
A second example is to use models \eqref{e:sige2} and \eqref{e:sigi2} and consider only inelastic dilatation ($\hat a_{\alpha\beta} = J_\inel\,A_{\alpha\beta}$) according to Sec.~\ref{s:indil}.
Contracting \eqref{e:evol0} with $a_{\alpha\beta}$ and using the relations from Sec.~\ref{s:indil} thus yields the evolution law
\eqb{l}
\dot J_\inel = \ds\frac{K}{\lambda}\frac{J}{I_1}\bigg(\frac{J}{J_\inel}-\frac{J_\inel}{J}\bigg)
\label{e:evol2}\eqe
for $J_\inel$. \\
For a generalized viscoelastic solid, see Fig.~\ref{f:rheo}b, evolution law \eqref{e:evol0} needs to be solved for $\hat a^{\alpha\beta}_M$ within each Maxwell element $M$.
This defines the stress $\sig^{\alpha\beta}_M$ in each Maxwell element. 
The total stress is then the sum of the stresses in all elements, i.e. 
\eqb{l}
\sig^{\alpha\beta} = \sig_{\el0}^{\alpha\beta}\big(a^{\gamma\delta}\big) + \ds\sum_{M=1} \sig_M^{\alpha\beta}\big(\hat a^{\gamma\delta}_M\big)\,.
\eqe
This model contains the special cases of a single Maxwell element -- for $M=1$ and $\sig_{\el0}^{\alpha\beta}=0$ -- and the Kelvin model -- for $M=1$ and $\hat a^{\alpha\beta} = a^{\alpha\beta}$ (see Fig.~\ref{f:rheo}c).

\textbf{Remark 7:} Apart of \eqref{e:sigi1} and \eqref{e:sigi2}, also the slightly different choices $\sig^{\alpha\beta}_{(\inel)} = -\eta\,\dot{\hat a}^{\alpha\beta}$ and $\sig^{\alpha\beta}_{(\inel)} = \lambda \,\dot J_\inel\,\hat a^{\alpha\beta}$ are consistent with the second law.

\textbf{iii. Surface constraints:}
Constraints are important for various applications. 
A popular example is incompressibility, which is discussed in the following. 
\\
\textit{Elastic incompressibility} implies $J_\el\equiv1$. 
This conditions leads to an extra stress that can for example be captured by the Lagrange multiplier method.
According to this, the stress follows from the potential
\eqb{l}
\hat\Psi = q\,\big(J_\el-1\big)\,,
\label{e:Psi4}\eqe
where $q$ is the corresponding Lagrange multiplier.
The constraint stress, according to \eqref{e:consti2} and \eqref{d:Jel_Eel}, thus is
\eqb{l}
\sig^{\alpha\beta}_{(\el)} = q\,a^{\alpha\beta}\,,
\label{e:sige4}\eqe
i.e.~it is dilatational.
The Lagrange multiplier $q$ is an additional unknown that needs to be solved for. 
It can be avoided by considering the penalty regularization
\eqb{l}
\hat\Psi = \ds\frac{K}{2}\big(J_\el-1\big)^2\,,
\label{e:Psi5}\eqe
where the in-plane bulk modulus $K$ is set to a very large value to ensure $J_\el\approx1$.
From \eqref{e:consti2} and \eqref{d:Jel_Eel} now follows
\eqb{l}
\sig^{\alpha\beta}_{(\el)} = K\big(J_\el-1\big)\,a^{\alpha\beta}\,.
\label{e:sige5}\eqe
\textit{Inelastic incompressibility} implies $J_\inel\equiv1$, i.e.~$\dot J_\inel=0$. 
According to \eqref{e:dJin2} this leads to the constraint
\eqb{l}
\dot{\hat a}^{\alpha\beta}\,\hat a_{\alpha\beta} = 0
\label{e:iic1}\eqe
on the internal variable $\hat a_{\alpha\beta}$.
This is a scalar equation, and so two more equations are needed in order to determine $\hat a_{\alpha\beta}$.
We can find those by contracting the evolution law with $\hat a_{\alpha\beta}$ and $a_{\alpha\beta}$.
For evolution law \eqref{e:evol1} we thus find
\eqb{l}
0 = \Lambda\,\big(1-J_\el^2\big) + G\,\big(2-4/I^\el_{1-}\big)
\label{e:iic2}\eqe
and
\eqb{l}
\dot{\hat a}^{\alpha\beta}a_{\alpha\beta} 
= \ds\frac{\Lambda}{\eta}\big(1-J_\el^2\big) + \frac{G}{\eta}\big(2-I^\el_1\big)\,.
\label{e:iic3}\eqe
Eqs.~\eqref{e:iic1}-\eqref{e:iic3} can then be solved for $\hat a_{\alpha\beta}$.

\textbf{iv. Surface plasticity:}
Plastic behavior can be characterized by the yield surface $f_\mry = f_\mry(\sig^{\alpha\beta}) = 0$ that satisfies $\dot f_\mry = 0$ during plastic flow.
Hence,
\eqb{l}
\ds\pa{f_\mry}{\sig^{\alpha\beta}}\,\dot\sig^{\alpha\beta} = 0\,.
\label{e:dfy}\eqe
A common approach to determine an evolution equation for $\hat a_{\alpha\beta}$ from this is to use the principle of maximum dissipation.
This assumes that for a given inelastic strain rate $\dot\eps^\inel_{\alpha\beta}$, the true stress is the one that maximizes the dissipation $\sig^{\alpha\beta}\dot\eps^\inel_{\alpha\beta}$ among all possible stress states.
This implies 
\eqb{l}
\dot\sig^{\alpha\beta}\dot\eps^\inel_{\alpha\beta} = 0\,.
\label{e:PMD}\eqe
Together, Eqs.~\eqref{e:dfy} and \eqref{e:PMD} imply that
\eqb{l}
\dot\eps^\inel_{\alpha\beta} = \ds\lambda\pa{f_\mry}{\sig^{\alpha\beta}}\,,
\label{e:evol3}\eqe
which is the evolution law for $\hat a_{\alpha\beta}$.
The scalar $\lambda$ follows from the condition $\dot f_\mry = 0$.
\begin{figure}[h]
\begin{center} \unitlength1cm
\begin{picture}(0,4.8)
\put(-7.1,-.2){\includegraphics[height=50mm]{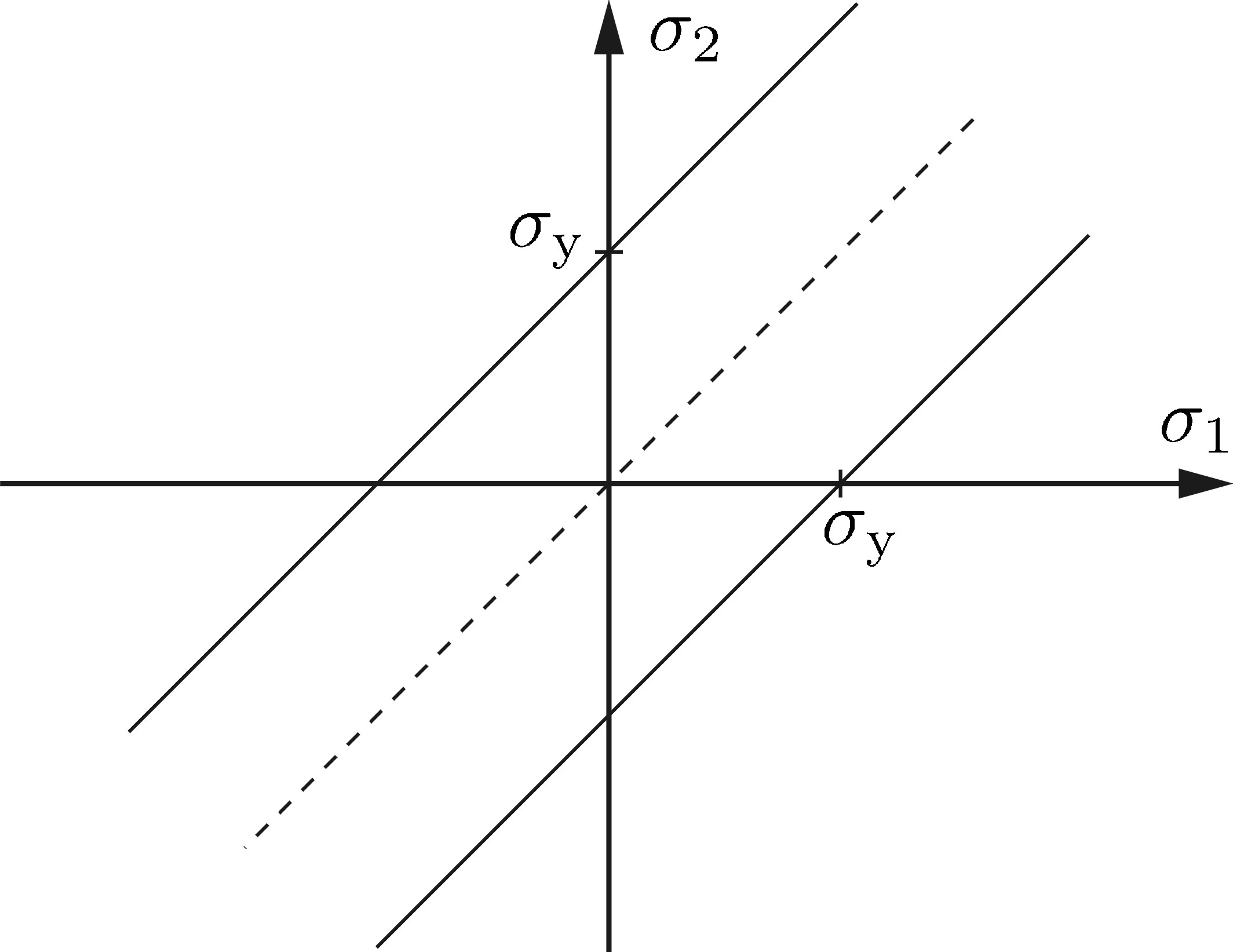}}
\put(0.3,-.2){\includegraphics[height=50mm]{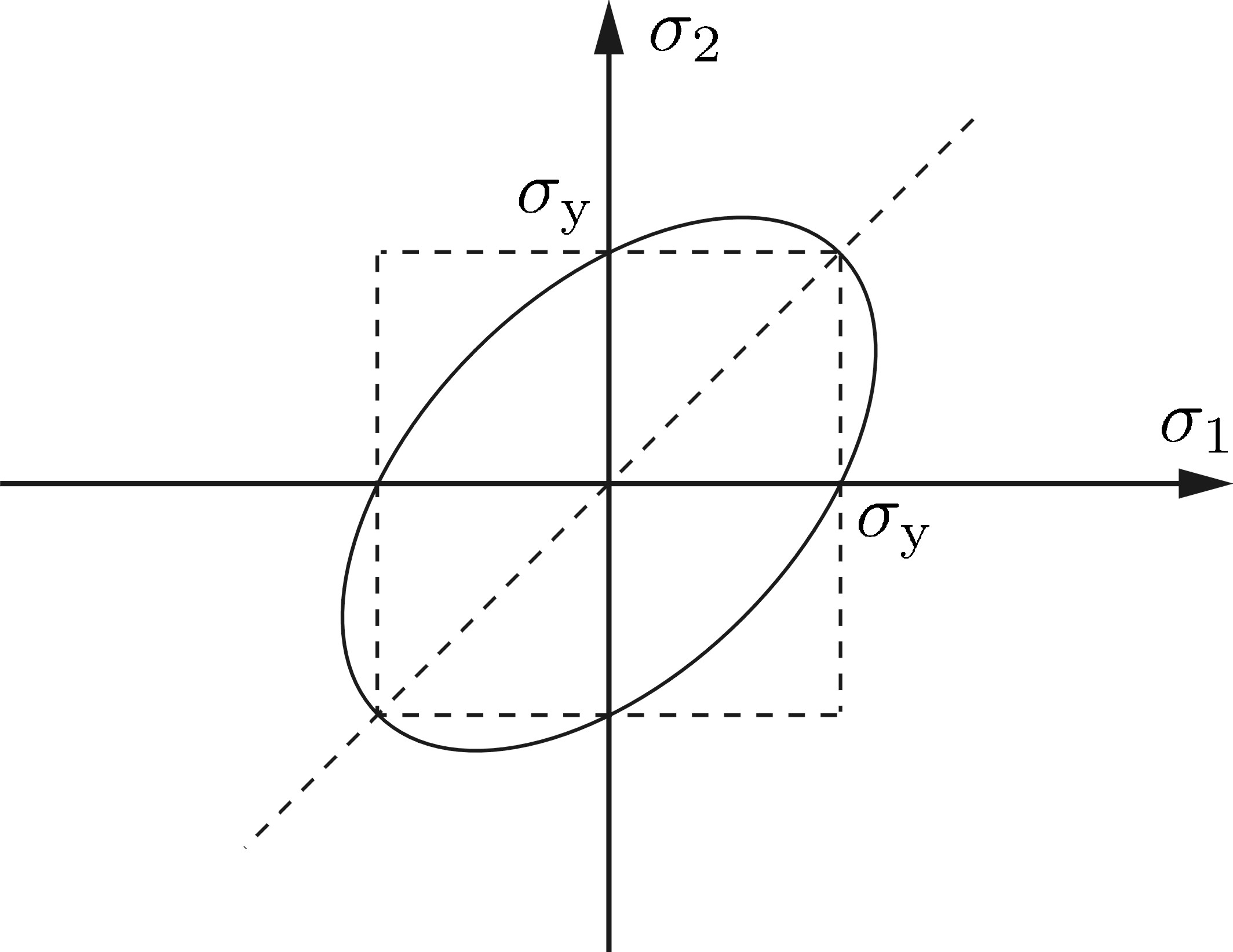}}
\put(-7.1,0){a.}
\put(0.3,0){b.}
\end{picture}
\caption{Surface plasticity: Yield surface $f_\mry=0$ in principal stress space for a.~2D von Mises plasticity and b.~3D von Mises plasticity.}
\label{f:plasti}
\end{center}
\end{figure}
\\
An example (that is a 2D version of von Mises plasticity\footnote{Another possibility is to use the classical 3D von Mises plasticity model together with the plane stress assumption. In this case $f_\mry := \norm{\tilde\bsig_\mathrm{dev}} - \sqrt{2/3}\,\sig_\mry$, where $\tilde\bsig_\mathrm{dev}$ is the full 3D stress deviator. See Fig.~\ref{f:plasti}b.}, see Fig.~\ref{f:plasti}a) is
\eqb{l}
f_\mry := \hat s - \ds\frac{\sqrt{2}}{2}\sig_\mry\,,
\label{e:fy}\eqe
where the material constant $\sig_\mry$ denotes the yield stress, and 
\eqb{l}
\hat s := \sqrt{\hat\sig^{\alpha\beta}_\mathrm{dev}\,\hat\sig_{\alpha\beta}^\mathrm{dev}}
\label{e:shat}\eqe
defines the 2D von Mises stress from the deviatoric stress 
\eqb{l}
\hat\sig^{\alpha\beta}_\mathrm{dev} := \hat\sig^{\alpha\beta} - \hat\gamma\,\hat a^{\alpha\beta}\,,\quad
\hat\sig_{\alpha\beta}^\mathrm{dev} := \hat a_{\alpha\gamma}\,\hat\sig^{\gamma\delta}_\mathrm{dev}\,\hat a_{\delta\beta} 
\eqe
and the surface tension
\eqb{l}
\hat\gamma := \frac{1}{2}\hat\sig^{\alpha\beta}\hat a_{\alpha\beta}\,,
\label{e:hatgam}\eqe
similar to definitions \eqref{e:gamma} and \eqref{e:Ndev}.
From
\eqb{l}
\ds\pa{...}{\sig^{\alpha\beta}} = J_\el\ds\pa{...}{\hat\sig^{\alpha\beta}}
\eqe
and
\eqb{l}
\hat\sig_{\alpha\beta}^\mathrm{dev}\,\hat a^{\alpha\beta} = \hat \sig^{\alpha\beta}_\mathrm{dev}\,\hat a_{\alpha\beta}=0\,,
\label{e:sig_dev_prop}\eqe
then follows
\eqb{l}
\ds\pa{f_\mry}{\sig^{\alpha\beta}} = J_\el\frac{\hat\sig_{\alpha\beta}^\mathrm{dev}}{\hat s}\,,
\eqe
such that
\eqb{l}
\dot\eps^\inel_{\alpha\beta} = \hat\lambda\ds\frac{\hat\sig^\mathrm{dev}_{\alpha\beta}}{\hat s}\,,
\label{e:evol4}\eqe
with $\hat \lambda := J_\el\,\lambda$, and
\eqb{l}
\sig^{\alpha\beta}\dot\eps^\inel_{\alpha\beta} = \lambda\,\hat s\,.
\eqe
Eq.~\eqref{e:consti1a} is thus satisfied for $\lambda\geq0$.
From \eqref{e:deps2}, \eqref{e:sig_dev_prop} and \eqref{e:evol4} follows that $\dot{\hat a}_{\alpha\beta}\,\hat a^{\alpha\beta}=0$, i.e.~plasticity model \eqref{e:evol4} is inelastically incompressible.

\subsubsection{Mechanical bending models}

\textbf{i. Bending elasticity:}
An example for a linear bending model is
\eqb{l}
\hat\Psi = \ds\frac{1}{2}\kappa^\el_{\alpha\beta}\,\hat f^{\alpha\beta\gamma\delta}\,\kappa^\el_{\gamma\delta}\,,
\label{e:Psi6}\eqe
where $\hat f^{\alpha\beta\gamma\delta}$ are the components of a constant material tensor.
For shells made of a homogenous material, those are given by
\eqb{l}
\hat f^{\alpha\beta\gamma\delta} = \ds\frac{h_0^2}{12}\, \hat c^{\alpha\beta\gamma\delta}\,,
\eqe
where $h_0$ is the initial shell thickness and $\hat c^{\alpha\beta\gamma\delta}$ is given by \eqref{e:cabcd}.
This bending model can be derived from the 3D St.-Venant-Kirchhoff material model via thickness integration, e.g.~see \citet{solidshell} for the purely elastic case.
From \eqref{e:consti3} and \eqref{e:Psi6} follows
\eqb{l}
\hat M^{\alpha\beta}_{(\el)} = \hat f^{\alpha\beta\gamma\delta}\,\kappa^\el_{\gamma\delta}\,.
\label{e:Mel1}\eqe
An example for a nonlinear bending model is
\eqb{l}
\hat\Psi = J_\el\,\bigg(k_\mrm\,\big(H - \hat H\big)^2 + \ds\frac{k_\mrg}{2}\,\big(\kappa - \hat\kappa\big)^2\bigg)\,,
\eqe
where $k_\mrm$ and $k_\mrg$ are material constants.
The model is an adaption and modification of the bending model by \citet{helfrich73}.
It produces the bending moment components
\eqb{l}
M^{\alpha\beta}_{(\el)} = \ds k_\mrm\,\big(H - \hat H\big)\,a^{\alpha\beta} + k_\mrg\,\big(\kappa-\hat\kappa\,\big)\,\tilde b^{\alpha\beta}
\label{e:Mel2}\eqe
and stress components 
\eqb{l}
\sig^{\alpha\beta}_{(\el)} = \ds\frac{\hat\Psi}{J_\el}a^{\alpha\beta} - 2k_\mrm\,\big(H-\hat H\big)\,b^{\alpha\beta} - 2k_\mrm\,\kappa\,\big(\kappa-\hat\kappa\big)\,a^{\alpha\beta}\,,
\eqe
due to \eqref{e:dhH}-\eqref{e:dHk2}.
Here, $\tilde b^{\alpha\beta}:=2H\,a^{\alpha\beta} - b^{\alpha\beta}$.

\textbf{ii. Bending viscosity:}
An analogous model to \eqref{e:sigi1} is the bending viscosity model
\eqb{l}
\hat M_{\alpha\beta}^{(\inel)} = \eta_\mrb\,\dot{\hat b}_{\alpha\beta}\,,
\label{e:Min}\eqe
since it satisfies \eqref{e:consti1a} for an analogous proof as for \eqref{e:sigi1}.
Setting this equal to $\hat M^{(\el)}_{\alpha\beta}$ (i.e.~assuming a Maxwell model) then yields the evolution law for $\hat b_{\alpha\beta}$.
For example taking \eqref{e:Mel1} with $\hat f^{\alpha\beta\gamma\delta} = f_0\,\big(\hat a^{\alpha\gamma}\hat a^{\beta\delta} + \hat a^{\alpha\delta}\hat a^{\beta\gamma}\big)/2$ yields the simple linear evolution law
\eqb{l}
\dot{\hat b}_{\alpha\beta} = \ds\frac{f_0}{\eta_\mrb}\Big(b_{\alpha\beta} - \hat b_{\alpha\beta} \Big)\,.
\label{e:evol5}\eqe

\textbf{iii. Bending plasticity:} 
Accounting for bending, the yield surface from Sec.~\ref{s:exM}.iv needs to be extended to $f_y\big(\sig^{\alpha\beta},M^{\alpha\beta}\big)$, such that during plastic flow
\eqb{l}
\ds\pa{f_\mry}{\sig^{\alpha\beta}}\,\dot\sig^{\alpha\beta} + \pa{f_\mry}{M^{\alpha\beta}}\,\dot M^{\alpha\beta} = 0\,.
\label{e:dfy2}\eqe
Invoking again the principle of maximum dissipation, which assumes that for given inelastic strain rates $\dot\eps^\inel_{\alpha\beta}$ and $\dot\kappa^\inel_{\alpha\beta}$, the true stress and bending moment components are those that maximize the dissipation 
$\sig^{\alpha\beta}\dot\eps^\inel_{\alpha\beta} + M^{\alpha\beta}\dot\kappa^\inel_{\alpha\beta}$ among all possible stress states, we find
\eqb{l}
\dot\sig^{\alpha\beta}\dot\eps^\inel_{\alpha\beta} + \dot M^{\alpha\beta}\dot\kappa^\inel_{\alpha\beta} = 0\,.
\label{e:PMD2}\eqe
Multiplying Eq.~\eqref{e:dfy2} by $\lambda$ and subtracting it from \eqref{e:PMD2} then gives
\eqb{l}
\ds\bigg(\dot\eps^\inel_{\alpha\beta} - \lambda\pa{f_\mry}{\sig^{\alpha\beta}}\bigg)\dot\sig^{\alpha\beta} 
+ \bigg(\dot\kappa^\inel_{\alpha\beta} - \lambda\pa{f_\mry}{M^{\alpha\beta}}\bigg)\dot M^{\alpha\beta} = 0\,,
\eqe
Since this is true for all $\dot\sig^{\alpha\beta}$ and $\dot M^{\alpha\beta}$ we find the flow rules \eqref{e:evol3} and
\eqb{l}
\ds\dot\kappa^\inel_{\alpha\beta} = \lambda\pa{f_\mry}{M^{\alpha\beta}}\,,
\label{e:evol6}\eqe
which are the evolution laws for $\hat a_{\alpha\beta}$ and $\hat b_{\alpha\beta}$.
The scalar $\lambda$ again follows from the condition $\dot f_\mry = 0$.
\\
As an example we consider the simple extension of \eqref{e:fy},
\eqb{l}
f_\mry := \ds\frac{\hat s}{\sig_\mry} + \frac{\hat s_\mathrm{M}}{M_\mry} - \frac{\sqrt{2}}{2}\,,
\label{e:fy2}\eqe
where the material constant $M_\mry$ denotes the yield limit for bending, and 
\eqb{l}
\hat s_\mathrm{M} := \sqrt{\hat M^{\alpha\beta}_\mathrm{dev}\,\hat M_{\alpha\beta}^\mathrm{dev}}
\eqe
with
\eqb{l}
\hat M^{\alpha\beta}_\mathrm{dev} := \hat M^{\alpha\beta} - \hat\gamma_\mathrm{M}\,\hat a^{\alpha\beta}\,,\quad
\hat M_{\alpha\beta}^\mathrm{dev} := \hat a_{\alpha\gamma}\,\hat M^{\gamma\delta}_\mathrm{dev}\,\hat a_{\delta\beta} 
\eqe
and
\eqb{l}
\hat\gamma_\mathrm{M} := \frac{1}{2}\hat M^{\alpha\beta}\hat a_{\alpha\beta}\,,
\eqe
is defined analogous to \eqref{e:shat}-\eqref{e:hatgam}.
Due to this analogy, the flow rule for $\kappa^\inel_{\alpha\beta}$ follows in analogy to \eqref{e:evol4}, which is still valid here, as 
\eqb{l}
\dot\kappa^\inel_{\alpha\beta} = \hat\lambda\,\ds\frac{\hat M^\mathrm{dev}_{\alpha\beta}}{\hat s_\mathrm{M}}\,,
\label{e:evol7}\eqe
which also satisfies Eq.~\eqref{e:consti1a} for $\lambda\geq0$. \\
The elasto-plasticity model described by the constitutive equations in \eqref{e:consti2}, \eqref{e:evol3} and \eqref{e:evol6}
is equivalent to the model of \citet{simo92b}. 
However, in \citet{simo92b}, \eqref{e:consti2} is written in terms of $\Phi = \rho_0\,\psi$, the Helmholtz free energy per unit reference area (in the case of $h=0$).
Also, \citet{simo92b}, consider an alternative model to \eqref{e:fy2} that is based on \citet{shapiro61}.

\subsubsection{Thermal models}

\textbf{i. Thermal energy:} Considering 
\eqb{l}
\hat\Psi = \ds C_\mathrm{H} \bigg[(T-T_0) - T\ln\frac{T}{T_0}\bigg]\,,
\label{e:Psi_t}\eqe
where $C_\mathrm{H}$ is a material constant and $T_0$ is a constant reference temperature \citep{holzapfel}, gives the specific entropy
\eqb{l}
s = \ds\frac{C_\mathrm{H}}{\hat\rho}\,\ln\frac{T}{T_0}\,,
\eqe
due to \eqref{e:consti1a} and \eqref{e:Psihat}.
This energy does not generate stresses. 
Those only appear in response to mechanical deformations.
Due to \eqref{e:Psi_t}, the stored energy (per intermediate area) is $\hat\rho\,u\,= \hat\rho\,(\psi+Ts) = C_\mathrm{H}(T-T_0)$.

\textbf{ii. Surface heat conduction:} A simple surface conductivity model satisfying \eqref{e:consti1a} is Fourier's law
\eqb{l}
q^\alpha = -k\,a^{\alpha\beta}\,T_{;\beta}\,,
\label{e:Fourier}\eqe
where the constant $k$ is the surface heat conductivity.
Model \eqref{e:Fourier} is analogous to Fick's law \eqref{e:Fick}.

\textbf{iii. Thermal surface expansion:}
A simple linear model for isotropic thermal expansion (analogous to chemical swelling) is
\eqb{l}
J_\inel = \lambda_\inel^2\,,\quad \lambda_\inel = 1 + \alpha_\mrT\,(T-T_0)\,,
\label{ex:Jin3}\eqe
where the material constant $\alpha_\mrT$ denotes the coefficient of thermal expansion.
Without loss of generality, one can then use $\hat a_{\alpha\beta} = J_\inel\,A_{\alpha\beta}$ as discussed in Sec.~\ref{s:indil}.
Model \eqref{ex:Jin3} leads to an additional entropy contribution due to \eqref{e:consti1b}, analogous to the contribution in $\mu$ seen in \eqref{e:full-mu}.

\textbf{iv. Thermal bending:}
An example for temperature induced curvature increase is the isotopic bending model \eqref{e:hatk_dil} with the linear curvature increase
\eqb{l}
\bar\kappa_\inel = 1 + \alpha_\kappa\,(T-T_0) \,,
\label{ex:kin4}\eqe
analogous to \eqref{ex:kin2}. Here $\alpha_\kappa$ is a material constant.
This curvature increase could be caused by a one-sided thermal expansion.
Analogous to \eqref{ex:kin3}, one can also consider the model
\eqb{l}
\bar\kappa_\inel = 1 + \alpha_\kappa\,\big(T_{+}-T_{-}\big) \,,
\label{ex:kin5}\eqe
where $T_{+}$ and $T_{-}$ denote the top and bottom temperatures of surface $\sS$, respectively.
Those need to be defined in a suitable way, e.g.~by using two separate PDEs of type \eqref{e:PDE-u} for the top and bottom surface.
Note that, models \eqref{ex:kin4} and \eqref{ex:kin5} lead to an additional entropy contribution due to \eqref{e:consti1b}, analogous to the contribution in $\mu$ seen in \eqref{e:full-mu}.

\section{Conclusion}\label{s:conc}

This work presents a general nonlinear shell theory for coupled elastic and inelastic deformations, accounting for growth, swelling, plasticity, viscosity and thermal expansion.
The formulation is derived from the balance laws of mass, momentum, energy and entropy using a multiplicative split of the surface deformation gradient into elastic and inelastic contributions.
The general constitutive equations of this coupling are derived and illustrated by several examples.
Those generally require the derivatives of various kinematical quantities w.r.t.~the elastic and inelastic deformations.
\\
Although the present formulation is purely theoretical, it is suitable for computational analysis, for example within the finite element method.
There has been important recent progress on rotation-free finite elements (FE) in the framework of isogeometric analysis \citep{kiendl09}.
Such FE formulations allow for a very accurate yet efficient surface description that is particularly beneficial for an accurate representation of curvatures.
It can thus be expected that isogeometric shell FE formulations for coupled inelastic and elastic deformations would be very beneficial.
So far, it seems that only elasto-plasticity and isotropic thermoelasticity have been analyzed with multiplicatively split isogeometric shell FE \citep{ambati18,inverseshell2}.
But the authors are currently applying the present theory to extend the hyperelastic graphene FE model of \citet{graphene3} to anisotropic thermoelasticity \citep{graphene4},
and to study the growth of fluid films using the FE model of \citet{droplet} and \citet{surfactant}.

\bigskip

{\Large{\bf Acknowledgements}}

The authors are grateful to the German Research Foundation (DFG) for funding this research under grant GSC 111. 
The authors also like to thank Farshad Roohbakhshan for feedback on the manuscript.

\appendix

\bibliographystyle{apalike}
\bibliography{bibliography,Reza_bib}

\end{document}